\ifx\fiverm\undefined
        \newfont\fiverm{cmr5} 
\fi
\input prepictex
\input pictex
\input postpictex
\catcode`\@=11
\@input{picmore.tex}
\documentstyle[12pt]{article}
\newlength{\dinwidth}
\newlength{\dinmargin}
\newcommand{\ra}{\rightarrow}

\newcommand{\ba}{\begin{array}}
\newcommand{\ea}{\end{array}}
\newcommand{\be}{\begin{equation}}
\newcommand{\ee}{\end{equation}}
\newcommand{\bea}{\begin{eqnarray}}
\newcommand{\eea}{\end{eqnarray}}

\def\d{{\rm d}}
\def\T{{\rm T}}
\def\N{{\rm N}}

\newcommand{\nn}{\nonumber}
\newcommand{\eq}[1]{(\ref{#1})}

\newcommand{\NP}[1]{ {\it Nucl.~Phys.} {\bf #1}}
 
\newcommand{\PL}[1]{ {\it Phys.~Lett.} {\bf #1}}

\newcommand{\JMP}[1]{ {\it J. Math.~Phys.} {\bf #1}}

\newfont{\sgcal}{eufm9}
\newfont{\smcal}{eusm9}
\newfont{\smbf}{msbm9}
\newfont{\gcal}{eufm10}
\newfont{\mcal}{eusm10}
\newfont{\mbf}{msbm10}
\newfont{\Gcal}{eufm10 scaled\magstep1}
\newfont{\Mcal}{eusm10 scaled\magstep1}
\newfont{\Mbf}{msbm10 scaled\magstep1}
\newcommand{\der}{\partial}
\newcommand{\eps}{\varepsilon}

\begin{document}
\thispagestyle{empty}
\addtocounter{page}{-1}
\begin{flushright}
SNUST 01-0901\\
PUPT 2009\\
{\tt hep-th/0110066}\\
\end{flushright}
\vspace*{1cm}
\centerline{\Large \bf Anatomy of Two-Loop Effective Action}
\vspace*{0.3cm}
\centerline{\Large \bf in}
\vspace*{0.3cm}
\centerline{\Large \bf Noncommutative Field Theories~\footnote{
Work supported in part by the BK-21 Initiative in Physics 
(SNU-Project 2), the KOSEF Interdisciplinary Research Grant 
98-07-02-07-01-5, the KOSEF Leading Scientist Program, 
the KRF Grant 2001-015-DP0082, and the KOSEF Brain-Pool Program.}}
\vspace*{1.2cm} 
\centerline{\bf Youngjai Kiem ${}^{a,d}$, Sung-Soo Kim${}^a$, 
Soo-Jong Rey ${}^{b,c}$, Haru-Tada Sato ${}^a$}
\vspace*{0.8cm}
\centerline{\it BK21 Physics Research Division \& 
Institute of Basic Science}
\vspace*{0.27cm}
\centerline{\it Sungkyunkwan University, Suwon 440-746 \rm KOREA ${}^a$} 
\vspace*{0.45cm}
\centerline{\it School of Physics \& Center for Theoretical Physics}
\vspace*{0.27cm}
\centerline{\it Seoul National University, Seoul 151-747 \rm KOREA ${}^b$}
\vspace*{0.45cm}
\centerline{\it Theory Division, CERN, CH-1211 Genev\'e \rm SWITZERLAND ${}^c$}
\vspace*{0.45cm}
\centerline{\it Physics Department, Princeton University, Princeton NJ 
08544 \rm USA ${}^d$}
\vspace*{0.8cm}
\centerline{\tt ykiem, sskim, haru@newton.skku.ac.kr 
\hskip0.45cm sjrey@gravity.snu.ac.kr }
\vspace*{1.2cm}
\centerline{\bf abstract}
\vspace*{0.3cm}
We compute, at two-loop order, one-particle-irreducible Green functions and 
effective action in noncommutative $\lambda[\Phi^3]_\star$-theory for both 
planar (g=0, h=3) and nonplanar (g=1, h=1) contributions. 
We adopt worldline formulation of the Feynman diagrammatics so that relation 
to string theory diagrammatics is made transparent in the Seiberg-Witten limit.
We argue that the resulting two-loop effective action is expressible via open 
Wilson lines: one-particle-irreducible effective action is generating 
functional of connected diagrams for interacting open Wilson lines.  
\vspace*{0.7cm}
\begin{flushleft}
\end{flushleft}

\baselineskip=18pt
\newpage

\section{Introduction}
\setcounter{section}{1}
\setcounter{equation}{0}
\indent
A notable feature of noncommutative field theories \cite{cds}
is a sort of 
duality between the ultraviolet (UV) and the infrared (IR) scale physics 
--- so-called UV-IR mixing \cite{uvir, uvir2} --- 
a phenomenon in sharp contrast
to the commutative, local quantum field theories. The duality is particularly
interesting, as it is reminiscent of the well-known `s-t channel duality' 
present in string theories. There, infinite tower of point-particle states
of string spectrum organizes in a modular invariant manner, and the
`s-t channel duality' interchanges the UV dynamics of open/closed strings
with the IR dynamics of closed/open strings. In fact, it is now known that
(a class of) noncommutative field theories arise quite naturally in the
so-called Seiberg-Witten limit \cite{seibergwitten}
of open string dynamics in the background of
nonzero closed string two-form potential, $B_{\rm mn}$. 
A viable interpretation would be that \cite{sjrey}, in the Seiberg-Witten 
limit, the `s-t channel duality' of underlying open string theories is 
partially retained and transmuted into the UV-IR mixing of resulting 
noncommutative field theories. 

Intuitively, the UV-IR mixing is understood as the manifestation that, in a 
generic noncommutative field theory, low-energy spectrum includes, in addition 
to point-like ones, dipole-like excitations \cite{dipole, stringdipole}. 
The simplest illustration is provided by the `Mott exciton' --- 
electron-hole bound-state in metal under strong magnetic field \cite{sjrey}. 
Argument for existence of the dipoles is based solely on noncommutative 
geometry
\bea
\left[x^m, x^n \right]  = i \theta^{mn}
\label{ncgeom}
\eea 
and nothing else. These dipoles are induced by turning on the spacetime
noncommutativity, Eq.(\ref{ncgeom}). As such, its physical characteristics
ought to depend on the noncommutativity parameter $\theta^{mn}$, as is 
illusrated elegantly by the so-called dipole relation between dipole's 
electric dipole moment $\Delta x$ and center-of-mass momentum $P$:
\bea
\Delta x^m = \theta^{mn} P_n.
\label{dipolerelation}
\eea
The UV-IR mixing then follows immediately from Eq.(\ref{dipolerelation}) --
a given momentum scale $P$ is mapped to, in addition to point-like excitations
of characteristic scale $(\Delta x)_{\rm particle} \sim \hbar \slash P$, 
dipole-like excitations of characteristic scale $(\Delta x)_{\rm dipole}
\sim \theta P$. Evidently, $(\Delta x)_{\rm particle} \sim 1 \slash
(\Delta x)_{\rm dipole}$, viz. what one might refer as UV and IR scales are 
excitation-dependent statement. 

The above assertion implies that, in noncommutative field theories, the 
dipoles ought to be viewed as 
collective excitations, not as elementary excitations, caused by turning on
the noncommutativity, Eq.(\ref{ncgeom}). Then an interesting question would 
be whether
the collective excitations -- dipoles -- are identifiable within noncommutative
field theories as a sort of (a complete set of) interpolating composite 
operators. In particular, in view of the aforementioned universality, these
operators ought to be present in all noncommutative field theories, be
they with gauge symmetry or not, or with Poincar\'e invariance or not. 
A conjecture
has been put forward \cite{sjrey} that these operators are nothing but
open Wilson lines, originally identified in the context of noncommutative 
gauge theories and S-duality therein \cite{owl}. 
In \cite{ours1, ours2}, 
for noncommutative $\lambda[\Phi^3]_\star$ scalar field theory, 
the conjecture was confirmed at one-loop level --- 
{\sl scalar} open Wilson lines are 
the interpolating operators for noncommutative dipoles obeying the 
dipole relation Eq.(\ref{dipolerelation}), and sum up nonplanar part of the
effective action. 

In this paper, to substantiate the one-loop confirmation \cite{ours1, ours2}
of the conjecture \cite{sjrey}, we extend 
computation of the N-point, one-particle-irreducible Green functions 
and effective action thereof to two-loop order. 
We will be adopting the worldline formulation of 
the $\lambda [\Phi^3]_\star$ theory, extending the
formulation constructed at one-loop order \cite{ours2}. 
We work with the formulation, as it is 
particularly suited for detailed comparison with the Seiberg-Witten limit of 
string theory computations. Compared to the one-loop case, two-loop
(and all higher-loop) computations exhibit nontrivial feature of the 
effective action: Green functions are classifiable into planar and 
nonplanar parts. At two-loop, the two parts correspond,
in the string theory counterpart, to Riemann surfaces of genus-zero with
three holes and genus-one with one hole, respectively. This is the feature
that will lead eventually to \underline{interacting} open Wilson lines, 
both at on-shell and off-shell.     

This paper is organized as follows.  In Section~2, we recapitulate aspects
of two-loop worldline formulation for commutative $\lambda [\Phi^3]$-theory
relevant for noncommutative counterpart. 
In Section~3, we develop the worldline formulation of noncommutative 
$\lambda [\Phi^3]_\star$-theory. In Section~4, we compute the N-point, 
one-particle-irreducible Green functions as integrals over the 
(N$+$3)-dimensional moduli space of two-loop vacuum Feynman 
diagram with N marked 
points. We show explicitly that, in contrast to the commutative counterpart, 
the Green functions obtained from $\Phi$-field insertions on planar vacuum 
diagrams are markedly different from those on nonplanar vacuum 
diagrams. Some tedious computational details are relegated to Appendix A.  
In Appendix B, via the standard Feynman diagrammatics, we confirm 
${\cal O}(\theta^3)$ effects for nonplanar Green functions, an aspect 
markedly different from the planar ones.  
In section 5, we discuss briefly, upon summing over N-point Green 
functions, how the two-loop effective action is expressible entirely
in terms of the scalar open Wilson lines. 

Our notations are as follows. The $d$-dimensional spacetime is taken
Euclidean, with metric $g_{mn}$. We use the shorthand notations as
\bea
p \cdot k = p_m g^{mn} k_n, \qquad
p \wedge k = p_m \theta^{mn} k_n \qquad
p \circ k = p_m (-\theta^2)^{mn} k_n.
\nn
\eea
Via Weyl-Moyal correspondence, products between fields are represented
by Moyal's product:
\bea
A(x) \star B(y) = \exp \left( - {i \over 2} \theta^{mn}
\partial_m^x \partial_n^y \right) A(x) B(y).
\nn
\eea
\section{Two-loop effective action of $\lambda[\Phi^3]$-theory}
\label{sec2}
\setcounter{section}{2}
\setcounter{equation}{0}
\indent
We begin with recapitulating worldline formulation of commutative 
$\lambda\Phi^3$ 
theory~\cite{ssphi3,2sato,HTS}. Specifically, we will compute two-loop
part of the effective action, bearing in mind of extensions to 
noncommutative $\lambda [\Phi^3]_\star$-theory in Section~3. 
We utilize the background field method and decompose $\Phi \rightarrow \Phi_0 
+ \varphi$, where $\Phi_0$ is the background field and $\varphi$ represents 
the quantum fluctuations. The generating functional, from which the Feynman 
rules are read off, is given by 
\bea
Z [\Phi_0] =Z_0[\Phi_0] \int{\cal D}\varphi
\exp\left( -\int \d^d x \, \Bigl\{ 
{1\over2}\varphi(x) \left(-\der^2_x+m^2+\lambda \Phi_0 \right) \varphi (x)
+ {\lambda\over3!}\varphi^3\Bigr\}
\right) \ ,              
\label{generate}
\eea
where $Z_0$ denotes the classical (tree) part. 
The resulting two-loop effective action is given, in terms of 
one-particle-irreducible vacuum diagram, by
\bea
\Gamma_2[\Phi]
&= &{\lambda^2\over2\cdot(3!)^2}  
\int{\cal D}\varphi \d^d x_1 \d^d x_2 \varphi^3(x_1) \varphi^3(x_2)
\exp\left[-{1\over2}\int \d^d x \varphi (x) \left(-\der^2_x+m^2+ \lambda 
\Phi_0(x) \right) \varphi(x) \right] \nn \\
&=&{\lambda^2\over2\cdot3!} \int \d^d x_1 \d^d x_2 \left<
x_2 \left| \left(\der^2_x+m^2+ \lambda \Phi_0 \right)^{-1} \right|x_1\right>^3
, \label{wickgamma}
\eea
where the second equality is obtained by applying the Wick contractions. The 
propagator in the last expression of Eq.(\ref{wickgamma}) is expressible, 
via the worldline formulation, as:
\bea
\label{pathprop}
\left<x_2 \left|{1\over-\der^2_x+m^2+ \lambda \Phi_0(x)} \right|x_1\right>
&=&\int\limits_0^\infty \d \T \int_{\scriptstyle y(0)=x_1 
        \atop\scriptstyle y(\T)=x_2}{\cal D}y(\tau){\cal D}k(\tau)
e^{-\int\limits_0^\T \d \tau \left(k^2+m^2-ik\cdot \dot{y}+ \lambda \Phi_0(y)
\right)} \nn \\
&=&\int\limits_0^\infty \d\T {\cal N}(\T)\int_{\scriptstyle y(0)=x_1 
        \atop\scriptstyle y(\T)=x_2}{\cal D}y(t)
e^{-\int\limits_0^\T \d \tau
\left( {1\over4}{\dot y}^2+m^2+ \lambda \Phi_0(y)\right)} .
\eea
Here, ${\cal N}(\T)$ is the normalization factor coming from the 
$k$-integral and ${\dot y} \equiv dy (\tau) / d\tau $. The 
effective action Eq.\eq{wickgamma} is then rewritable, after Taylor-expanding
the background field interaction, in the following form: 
\bea
\Gamma_2 [\Phi_0] &=&{\lambda^2\over12} \sum_{\rm N_1,N_2,N_3=0}^\infty
{(- \lambda )^{\rm N_1+N_2+N_3}\over \rm N_1!N_2!N_3!} \int \d^d x_1
\d^d x_2 \prod_{a=1}^3\int_0^\infty \d \T_a e^{-m^2 \T_a}{\cal N}(\T_a) 
\label{Gammaplane}\\
&\times&\int_{\scriptstyle y_a(0)=x_1 
        \atop\scriptstyle y_a(\T_a)=x_2}{\cal D}y_a \left(\tau^{(a)} \right)
\exp\left( \,-\int_0^{\T_a}\d \tau^{(a)} 
{1\over4}\dot{y}^2_a \left(\tau^{(a)} \right) \,\right)
\left[\,\int_0^{\T_a}\d  \tau^{(a)} \,
\Phi_0 \left(y_a(\tau^{(a)}) \right) \,\right]^{{\rm N}_a}. \nn
\eea
The three Feynman-Schwinger parameters, $\T_a$ ($a=1,2,3$), specify moduli
of the two-loop vacuum diagram.  The number of background $\Phi_0$
lines attached to the $a$-th internal propagator is denoted as $\N_a$.  

The $\N$-point Green functions are extracted from Eq.\eq{Gammaplane} by 
replacing the background $\Phi_0$'s with monochromatic plane-waves 
$\Phi_0(y(\tau))$ $\ra$ $\sum_{j=1}^{\rm N} \exp[ip_j \cdot y(\tau)]$.
The $\delta$-functions imposing energy-momentum conservation at the two
interaction vertices follow evidently from integration over $x_1, x_2$.
Because of them, terms containing the same energy-momentum vector
more than once can be discarded.
Introduce `scalar vertex operator':
\bea\label{vertex}
V^{(a)} (p) := \int_0^{\T_a}\d\tau^{(a)}\exp\left[ip \cdot y_a
\left(\tau^{(a)}\right)\right]\ ,
\nonumber
\eea
and replace the background $\Phi_0$'s into monochromatic plane-waves:
\bea
{1\over {\rm N}_a!}\left[\int_0^{\T_a}
\Phi_0 \left(y_a(\tau^{(a)}) \right)\d\tau^{(a)}\right]^{{\rm N}_a}
\longrightarrow \sum_{i_1<i_2<\cdots<i_{{\rm N}_a}} 
V^{(a)}_{i_1} \cdots V^{(a)}_{i_{{\rm N}_a}}\ , \label{replaceV}
\nonumber
\eea
where each $i_k$ ($k=1,\dots,\N_a$) runs from 1 to $\N$ in so far as the 
time-ordering is obeyed.  We then find, from Eq.(\ref{Gammaplane}),
the two-loop part of the one-particle-irreducible, $\N$-point Green function: 
\bea
\Gamma_\N[\Phi_0] 
&=&{1 \over 12} (-\lambda)^{\N+2} \sum_{\N_1,\N_2,\N_3=0}^\N
\sum_{\sigma(\N_1,\N_2,\N_3)}\int \d^dx_1 \d^d x_2 \prod_{a=1}^3\int_0^\infty 
\d\T_a e^{-m^2\T_a}{\cal N}(\T_a)      \nn \\
&\times&\int_{\scriptstyle y_a(0)=x_1 
        \atop\scriptstyle y_a(\T_a)=x_2}{\cal D}y_a(\tau^{(a)})
\exp\left[-\int_0^{\T_a} \d\tau^{(a)} {1\over4}\dot{y}^2_a \right]
\prod_{n=1}^{\N_a} \int_0^{\T_a} \d\tau_n^{(a)}
e^{ip_n^{(a)} \cdot y(\tau_n^{(a)})}. 
\label{Gammaplanenew}  
\eea
Here, $\N_1 + \N_2 + \N_3 = \N$, and 
$\sigma(\N_1, \N_2, \N_3)$ refers to all possible graph-theoretic
combinatorics for attaching $\N_a$ momenta to each of the three 
internal propagators, irrespective of ordering of the external 
$\Phi_0$ lines. Expand $y_a(\tau)$ into normal modes
\bea
y_a(\tau)=x_1+{\tau\over \T_a}(x_2-x_1)+\sum_{m=1}^\infty
y_m \, {\rm sin}\Bigl({m\pi\tau\over \T_a}\Bigr) \,
\nonumber
\eea
and perform the functional integral over $y_a$. 
Integration over $y_m$ ($m = 1, 2, \cdots)$ yields
\bea
{\cal N}(\T) 
\int_{\scriptstyle y(0)=x_1 
        \atop\scriptstyle y(\T)=x_2}{\cal D}y(\tau)\exp\left[\, 
-\int_0^\T \d\tau {1\over4}{\dot y}^2(\tau) \,\right]
=\left({1\over4\pi \T}\right)^{d\over2}
\exp\Biggl[\,-{(x_1-x_2)^2\over4 \T}\,\Biggr], 
\nonumber
\eea
while subsequent integration over $x_i$ ($i=1,2$) yields the N-point Green
function:
\bea
\Gamma_\N={1\over12}\sum_{\{\N_a\}=0}^\N
\sum_{\{\sigma\}} \,
(2\pi)^d \delta\left(\sum_{a=1}^3\sum_{n=1}^{\N_a}p_n^{(a)}\right)
\,\Gamma^{(\N_1,\N_2,\N_3)} 
\nonumber
\eea
with
\bea
\Gamma^{(\N_1,\N_2,\N_3)}&=& {(-\lambda)^{\N+2}\over(4\pi)^d}
 \prod_{a=1}^3 \int_0^\infty \d\T_a e^{-m^2 \T_a} 
\Delta^{d \over 2}(\T) \int_0^{\T_a} 
\prod_{n=1}^{\N_a} d\tau^{(a)}_n  \label{masterS} \\ 
&\times& \!\!\! \exp \Big[ {1\over2}\sum_{a=1}^3 \sum_{j,k}^{\N_a} 
p^{(a)}_j G^{\rm sym}_{aa} \left(\tau^{(a)}_j,\tau^{(a)}_k\right) p^{(a)}_k
 +\sum_{a=1}^3\sum_{j=1}^{\N_a}\sum_{k=1}^{\N_{a+1}} 
p^{(a)}_jG^{\rm sym}_{aa+1}\left(\tau^{(a)}_j,\tau^{(a+1)}_k\right)p^{(a+1)}_k 
             \Big] . 
\nonumber
\eea
Here, $\Delta(\T)$ denotes the following combination of the two-loop
vacuum diagram moduli parameters:
\bea
\Delta(\T) = \left(\T_1 \T_2+\T_2 \T_3+\T_3 \T_1 \right)^{-1} \ .
\nonumber
\eea
Note that, in Eq.(\ref{master2}), 
the Green function is summarized in terms of the worldline 
correlators~\cite{ssphi3}:
\bea
G^{\rm sym}_{aa}(x,y)&=&|x-y|-\Delta(\T) (\T_{a+1}+\T_{a+2})(x-y)^2\ , 
\label{Gsaa}\\
G^{\rm sym}_{aa+1}(x,y)&=&(x+y)-\Delta(\T) 
\left(x^2 \T_{a+1}+y^2 \T_a+(x+y)^2 \T_{a+2}\right)\ . 
\label{Gs12}
\eea
The worldline correlators Eqs.(\ref{Gsaa}, \ref{Gs12}) coincide precisely with
the string theory worldsheet correlators \cite{HTS,add} in the zero-slope
limit, $\alpha' \rightarrow 0$.  Likewise, the moduli-dependent measure factors
$ \left(\prod_{a=1}^3 e^{-m^2 \T_a}\right) \Delta^{d \over 2}(\T) $ 
originate from the integrand of the string worldsheet partition function
in the same limit.
\section{Two-Loop Effective Action of $\lambda [\Phi^3]_\star$-theory}
\setcounter{section}{3}
\setcounter{equation}{0}
\indent
We now extend worldline formalism of the $\lambda [\Phi^3]_\star$-theory 
investigated in \cite{ours2}. Again, expand the action around a 
background field: $\Phi = \Phi_0 + \varphi$. After Wick rotation to the 
Euclidean space, it reads 
\bea
S&=&S^{\rm 1-loop} + S_{\rm int}\nn\\
&=&
\int \d^d x\left(\, {1\over2}(\der\varphi)^2 +{1\over2}m^2\varphi^2 + 
{\lambda \over2}\varphi\star \Phi_0\star\varphi\,\right)(x) + 
\int \d^d x\, \left( {\lambda 
\over3!}\varphi\star\varphi\star\varphi \,\right)(x)\ .
\nonumber
\eea
Taking the momentum-space representation, with which the one-loop 
effective action computation has become simplified enormously 
\cite{ours1}, we obtain:
\bea
S^{\rm 1-loop}&=&{1 \over 2} 
\int{\d^d k_1\over(2\pi)^d}{\d^d k_2\over(2\pi)^d}
       \widetilde{\varphi}(k_1)\widetilde{\varphi}(k_2)
       \left[ (k^2+m^2) (2 \pi)^d \delta^{(d)}(k_1 + k_2)\right.\nn\\
&&+{\lambda\over2}\int{\d^d p\over(2\pi)^d} (2\pi)^d \delta^{(d)}(p+k_1+k_2)
   \left(e^{{i\over2}k_1 \wedge p} +e^{-{i\over2}k_1\wedge p}\right)
   \left.\widetilde{\Phi_0}(p)\right] 
\nn \\
\label{S1}\\
S_{\rm int}&=&{\lambda \over3!}\int \prod_{a=1}^{3}\frac{\d^d k_a}{(2\pi)^d}
   (2\pi)^d\delta^{(d)} \left(\sum_{a=1}^3 k_a \right) e^{{i\over2}{k_1 \wedge k_2}}
   \widetilde{\varphi}(k_1)\widetilde{\varphi}(k_2)\widetilde{\varphi}(k_3)
.
\label{Sint}
\eea
Schematically, the two-loop effective action is $\star$-product deformation
of the commutative effective action, Eq.\eq{wickgamma}: 
\bea\label{2effa}
\Gamma^{\rm 2-loop}[\Phi_0]
&=&\int{\cal D}\varphi\,  e^{-S_{\rm 1-loop}}\,
\frac{i^2}{2!}(S_{\rm int})^2  \nn \\
&=& {\lambda^2\over 2\cdot (3!)^2} \int{\d^d x_1}{\d^d x_2}\, 
\Bigl\langle\varphi^3_\star (x_1)
\varphi^3_\star(x_2) \Bigr\rangle_{\Phi_0} \ ,
\eea
where $\langle\,\,\,\,\rangle_{\Phi_0}$ 
denotes Wick contraction by the background-dependent propagator.
Thus, the two-loop vacuum diagram is specified entirely by $S_{\rm int}$, 
while insertion of external $\Phi_0$-background interactions is governed 
by $S_{\rm 1-loop}$. 
Using the momentum-space representation Eq.\eq{Sint}, we obtain 
\bea
\Gamma^{\rm 2-loop}[\Phi_0]
&=&{\lambda^2\over 2\cdot (3!)^2}
\int \prod_{a=1}^{3}\frac{\d^d k_a}{(2\pi)^d}
   (2\pi)^d\delta^{(d)} \left(\sum k_a \right)\, 
   e^{\frac{i}{2}{k_1 \wedge k_2}} \nn \\
&&\times \int \prod_{b=1}^{3}\frac{\d^d l_b}{(2\pi)^d}
   (2\pi)^d \delta^{(d)} \left(\sum l_b \right) \,
   e^{\frac{i}{2}{l_1 \wedge l_2}}
\Bigl\langle 
{\widetilde\varphi}(k_1){\widetilde\varphi}(k_2){\widetilde\varphi}(k_3)
{\widetilde\varphi}(l_1){\widetilde\varphi}(l_2){\widetilde\varphi}(l_3)
\Bigr\rangle_{\Phi_0} .
\nn
\eea
Begin with two-loop vacuum diagrams, for which we set $\Phi_0 = 0$.
Thus, the propagator is simply given by   
\bea\label{Wickf}
\Big\langle\widetilde{\varphi}(k_1)\widetilde{\varphi}(l_1)\Big\rangle_0
= (2\pi)^d \delta^{(d)}(k_1 +l_1)\frac{1}{(k_1^2+m^2)} .
\eea
Working out the phase-space integrals for each of the six possible
combinatorics (neglecting tadpoles), we obtain, after $\ell_a$-integrals,
\bea
\Gamma^{\rm 2-loop}[0]&=&{3 \lambda^2\over 2\cdot (3!)^2}
\int \d^d x_1 \d^d x_2 
\int \prod_{a=1}^{3}\frac{\d^d k_a}{(2\pi)^d}
\frac{e^{ik_a \cdot (x_1-x_2)}}{(k_a^2+m^2)}  \,
\Bigl(\, 1+e^{ik_1\wedge k_2}\,\Bigr)\
\nn \\
&:=& \Big( {\bf P} + {\bf NP} \Big),
\nn
\eea
viz. a sum of two types of vacuum diagrams, ${\bf P}$ and ${\bf NP}$.
Diagrammatically, the phase-factor $e^{ik_1\wedge k_2}$ in ${\bf NP}$ 
originates from crossing~\cite{uvir} two of the three internal propagators, 
as is evident from the noncommutative Feynman diagramatics.  
We refer the two diagrams ${\bf P}$ and ${\bf NP}$ as \underline{planar} and 
\underline{nonplanar} vacuum diagrams, respectively. They are depicted in 
{}~Fig.~1. As is evident from `thickening' the internal propagators, they
are mapped, at zero slope limit, to the following string worldsheet diagrams:
\hfill\break
\vskip0.1cm
\noindent
$\bullet$ \underline{planar vacuum diagram} $\longrightarrow$ sphere with 
three holes (g=0, h=3), \hfill\break
$\bullet$ 
\underline{nonplanar vacuum diagram} $\longrightarrow$  torus with a hole
(g=1, h=1).  
\hfill\break
\vskip0.1cm
\noindent

Actually, the two types of the vacuum Feynman 
diagrams are related each other. In the 
configuration-space Feynman diagrammatics, the nonplanar vacuum diagram 
${\bf NP}$ is obtainable from the planar one ${\bf P}$ by inserting 
Moyal's pseudo-differential phase-operator: 
\bea \label{vac}
\Gamma^{\rm 2-loop}[0]  & = & {\lambda^2\over 24} 
\int \d^d x_1 \d^d x_2 (1+e^{-i \partial_z \wedge \partial_w}) \\
 & & \times  
 \left. \int \prod_{a=1}^{3}\frac{\d^d k_a}{(2\pi)^d} \,
   \frac{e^{ik_1 \cdot (x_1-z)}e^{ik_2 \cdot (x_1-w)}e^{ik_3 \cdot (x_1-x_2)}}
{(k_1^2+m^2)(k_2^2+m^2)(k_3^2+m^2)}\right|_{z=w=x_2}\ . \nn
\eea

Having obtained the two-loop vacuum diagrams for $\Phi_0 = 0$, we readily
obtain the vacuum diagrams for nonzero $\Phi_0$-background by replacing the 
free propagators in Eq.\eq{vac} into the full, background-dependent propagator.

\vspace{5mm}
\begin{minipage}[htb]{15cm} 
\begin{center}
\input{fig1.pictex}
\end{center}
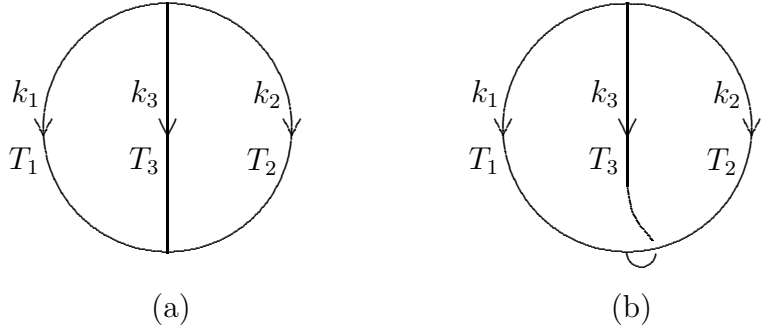
{\bf Figure 1:} {\sl Two types of two-loop vacuum diagram in $\lambda
[\Phi^3]_\star$ field 
theory are shown. The diagram (a) is a planar vacuum bubble and 
the diagram (b) is a nonplanar vacuum bubble.} 
\end{minipage}
\vspace{8mm}

\noindent
Hence, we will first derive noncommutative counterpart of the scalar 
propagator \footnote{
Our phase-factor convention is different from the analysis given in 
\cite{ours1, ours2}. 
The different conventions, however, affect intermediate steps
only, and yield identical final result.} .
Inferring from  
\bea
\int \d^d x \varphi(x)\star\Phi_0(x)\star\varphi(x) =
\int {\d^d k_1 \over (2 \pi)^d} {\d^d k_2 \over (2 \pi)^d}
{\d^d p \over (2 \pi)^d}
(2 \pi)^d \delta^d (k_1+k_2+p)e^{-{i\over2}k_1\wedge p}
{\widetilde \varphi}(k_1){\widetilde\Phi_0}(p){\widetilde\varphi}(k_2)\ ,
\nonumber
\eea
when obtaining the background-dependent scalar propagator by taking 
functional derivative with respect to $\varphi$,
a natural prescription of Moyal's phase-factor would be to strip off the 
$e^{-{i\over2}k_1\wedge p}$ factor from the interaction term in Eq.$\eq{S1}$. 
Implicit to this prescription is that, upon expanded in powers of the 
background field, products of a string of $\widetilde{\Phi_0}$'s is interpreted
in terms of the $\star$-product. This will be elaborated further in the 
next section. We have verified consistency of this prescription for one-loop 
computation by comparing the result with that in \cite{ours1, ours2}, 
and hence expect the consistency for two-loop computation as well. 
Being so, we express the background-dependent scalar propagator 
(noncommutative counterpart of Eq.\eq{pathprop}) in the worldline formalism 
as  
\bea
\Delta_a(x_1, x_2)&=& 
\left< x_2 \left\vert {1 \over - \partial_x^2 +m^2 + { \lambda \over 2}
\Phi_0(x) \star } \right\vert x_1 \right>
\nonumber \\
&=&
\int\limits^\infty_0 \d\T_{x_1} 
\int_{\scriptstyle y_a(0)={x_2} 
        \atop\scriptstyle y_a(\T_a)=w}
{\cal D}y_a(\tau^{(a)})
{\cal D}k_a(\tau^{(a)})  \label{Da} \nn \\
&\times& \!\!\!\! {\cal P}_{\rm t} \exp
\left[ -\int^{\T_a}_0 \d\tau^{(a)} \left( k^2_a+m^2 - ik_a\dot{y}_a 
+{\lambda \over2} \int \frac{\d^d p}{(2\pi)^d}
(1 + e^{i k_a\wedge p})\widetilde{\Phi_0}(p)
\right)\right]. 
\label{GO}
\eea

The two types of interaction vertices, $(1 + e^{i k_a\wedge p})$, 
represent, in the 
double-line notation of the noncommutative Feynman diagrams, insertions on 
the `left' and `right' side (with crossing), respectively.  
One can verify that the one-loop effective action computed in
\cite{ours1, ours2} is readily derived by imposing periodic boundary condition 
($x_1 = x_2$) and taking into account of Jacobian associated with Killing
symmetry, $1/\T_a$: $\d \T_a \rightarrow {\d\T_a \slash \T_a}$.
The two-loop effective action is now obtained by Wick-contracting Eq.(3.3)
and is expressible schematically as a sum of the planar and the nonplanar
parts:
\bea
\Gamma^{\rm 2-loop}[\Phi_0]=\Gamma^{\bf P}[\Phi_0] + 
\Gamma^{\bf NP}[\Phi_0]\,
\nn
\eea
where
\bea
\label{GP}
\Gamma^{\bf P}[\Phi_0] &=& {\lambda^2\over 24}
\int \d^d x_1 \d^d x_2 \, 
\Delta_1(x_2,x_1) \star \Delta_2(x_2,x_1) \star \Delta_3(x_2,x_1)\ ,
\\
\label{GNP}
\Gamma^{\bf NP}[\Phi_0] &=& {\lambda^2\over 24}
\int \d^d x_1 \d^d x_2 \, e^{-i \partial_z \wedge \partial_w}
\Delta_1(z,x_1) \star \Delta_2(w,x_1) \star \Delta_3(x_2,x_1)
\Bigr|_{z=w=x_2}\ .
\eea
The $\star$-product between the propagators summarizes the Filk phase-factors
\cite{filk}
associated with the {\sl vacuum} Feynman diagram. The additional insertion
of the pseudo-differential operator, $e^{ - i \partial_z \wedge \partial_w}$,
for the nonplanar part accounts for the extra phase-factor associated with
the nonplanar crossing. See Fig. 1(b). 

We end this section with a cautionary remark concerning our nomenclature.
The \underline{planar} part Eq.(\ref{GP}) of the effective action 
summarizes one-particle-irreducible Green functions arising from
{\sl planar} and {\sl nonplanar} insertions of the background $\Phi_0$-fields 
on the {\em planar} vacuum diagram. Likewise, the \underline{nonplanar}
part Eq.(\ref{GNP}) of the effective action summarizes one-particle-irreducible
Green functions arising from arbitrary insertions of the background 
$\Phi_0$-fields on the {\em nonplanar} vacuum diagram 
\footnote{While notationally
consistent, we trust this will be a source of profound confusion to the 
readers.}.  
\section{The $\N$-point One-Particle-Irreducible Green functions}
\setcounter{section}{4}
\setcounter{equation}{0}
\indent 
In this section, we will compute the N-point, one-particle-irreducible 
Green functions, and express them as (N$+$3)-dimensional integral over 
the moduli space of two-loop Feynman diagram with N marked points. 
The expression will eventually enable us to resum them over $\N$, and 
obtain the two-loop effective action \footnote{
Because of variety of technical complications, the resummation and
computation of the two-loop effective action will be relegated to a 
separate paper \cite{pants}.}. In computing the Green functions, we 
will pave essentially the same steps as Eq.\eq{Gammaplane} through 
Eq.\eq{Gammaplanenew} in section 2, taking, at the same time, 
proper care of Filk's phase-factors resulting from insertions of 
the background $\Phi_0$ field on each internal propagator. 

\vspace{0mm}
\begin{minipage}[htb]{15cm} 
\begin{center}
\input{fig2.pictex}
\end{center}
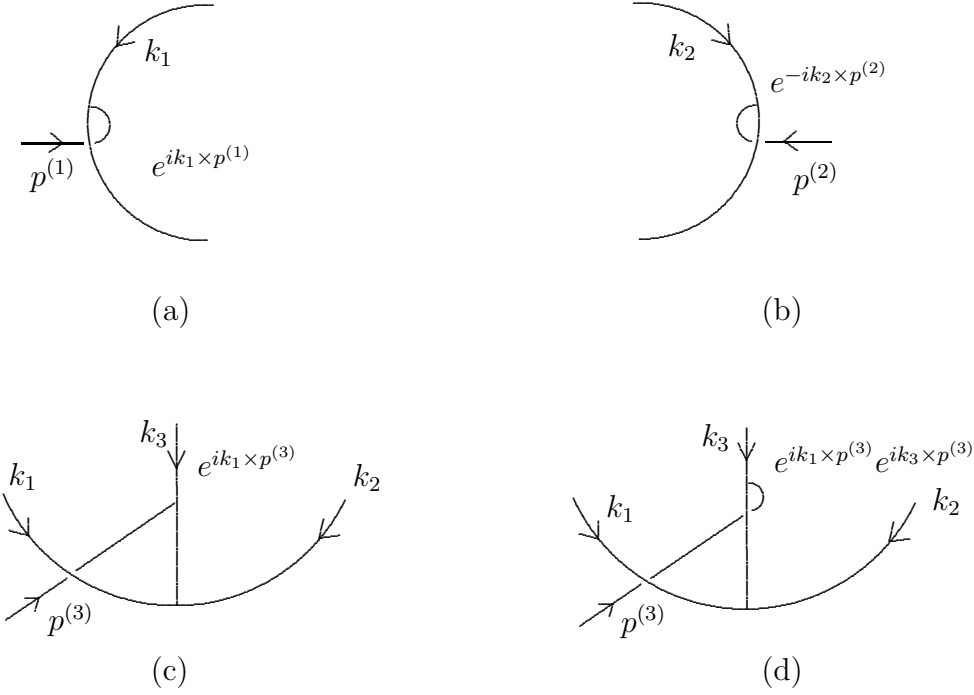
{\bf Figure 2:} {\sl Assignments of phase factors resulting from the
crossing of external insertions and internal lines.}
\end{minipage}
\vspace{8mm}

\subsection{Moyal's Phase-Factors}
As Filk's phase-factors are the new ingredients in the noncommutative Feynman
diagrammatics, we will begin with enumerating and fixing convention of them. 
Apart from insertion of the pseudo-differential operator, 
$e^{-i \partial_z \wedge \partial_w}$, the expressions Eqs.(\ref{GP}, 
\ref{GNP}) of $\Gamma^{\bf P}$ and $\Gamma^{\bf NP}$ comprise the same 
integrand -- triple product of the $\Phi_0$-field dependent propagators.  
To extract Moyal's phase-factors, we first Taylor-expand each propagator 
in powers of $\Phi_0$'s, precisely as in Eq.\eq{Gammaplane}. 
By doing so, we readily observe that two types of interaction vertices are 
generated: $1 \cdot \widetilde{\Phi_0}$ for the planar $\Phi_0$-insertion, and 
$e^{ik_a\wedge p} \widetilde{\Phi_0}$ for the nonplanar $\Phi_0$-insertion.
One might thus suppose that all the phase-factors are prescribable by 
introducing an insertion-specific phase-factor:  
\bea
\label{N1}
\exp[i k_a \wedge p^{(a)}_j \eps^{(a)}_j]\ , \qquad j=1,2,\cdots,\N_a\
\eea 
where $\eps^{(a)}_j = 0$ and 1 for planar and nonplanar $\Phi_0$-insertions, 
respectively. It turned out that this prescription is valid if and only if
there is only one independent internal momentum, viz. one-loop diagram. 
For two-loop diagrams, a closer inspection of the diagrammatics depicted 
in {}~Fig.~2 is imperative. 

We will assign the Filk phase-factors beginning with $(a=1)$ internal 
propagator and sequentially with $(a=2)$, $(a=3)$ ones. 
For the $(a=1)$ internal propagator, the 
phase-factor assignment Eq.(\ref{N1}) is complete (Fig.~2(a)). 
Next, for the $(a=2)$ internal propagator, as (Fig.~2(b)) indicates,
a $\Phi_0$-field insertion along the `left' edge crosses 
the $(a=2)$ internal propagator carrying momentum $k_2$. Because of 
the crossing, the insertion-specific phase-factor, Eq.\eq{N1}, ought to be
assigned with $\eps^{(2)}_j=0$ and $-1$ for planar and nonplanar insertions,
respectively. 
Lastly, for the $(a=3)$ internal propagator, we will prescribe the
phase-factor convention as follows. Let the insertions to the `right' edge
to cross the $(a=3)$ internal propgator (just as the insertions to the
`right' edge for the $(a=1)$ internal propagator), and assign the 
insertion-specific phase-factor, Eq.(\ref{N1}), with $\eps^{(3)} = 0$ and $1$ 
for the `left' and the `right' insertions, respectively. Also, let these 
insertions eventually cross the $(a=1)$ internal propagator --- a convention
giving rise to an extra phase-factor 
$e^{ik_1\wedge p_j^{(3)}}$; $j=1,2,\cdots,\N_3$ for both 
`left' and `right' insertions along the $(a=3)$ internal propgator. See 
{}~Figs.~2(c) and 2(d). 
The resulting phase-factor assignment 
\footnote{Different crossing prescriptions may be adpoted
for the $\Phi_0$-field insertions, but they all turn out equivalent 
once the energy-momentum conservation conditions at each vertex are 
imposed.} of $\eps^{(a)}$'s is depicted in {}~Fig.~3(a).

\vspace{8mm}
\begin{minipage}[htb]{15cm} 
\begin{center}
\input{fig3.pictex}
\end{center}
{\bf Figure 3:} 
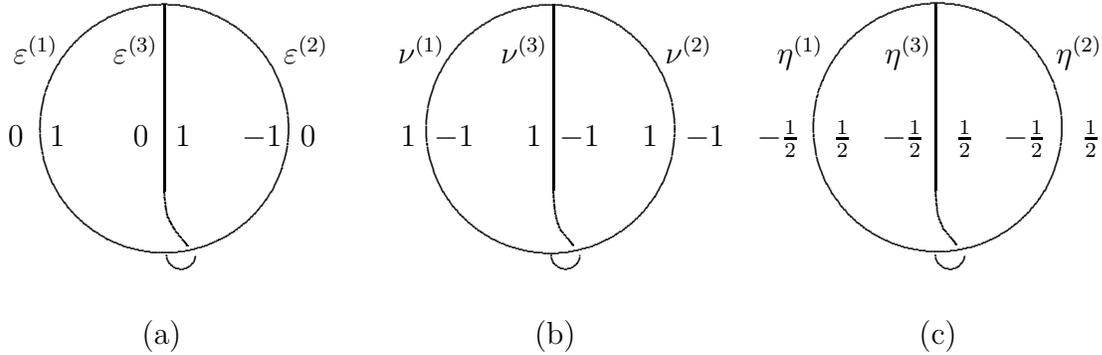
{\sl The assignments of $\eps^{(a)}$, $\nu^{(a)}$ and $\eta^{(a)}$ 
for six possible types of external insertions.  The assignments 
are identical for the planar vacuum diagrams.}
\end{minipage}
\vspace{8mm}

Collecting all the phase-factors arising from the Taylor expansion, 
 the term with $\N_1, \N_2, \N_3$ powers of the $\Phi_0$-field attached to
$a=1,2,3$ internal propagators, respectively, is accompanied by
\bea
\label{addphase}
\left(\prod_{j=1}^{\N_1}e^{i{k_1}\wedge p^{(1)}_j \eps^{(1)}_j}\right)
\left(\prod_{j=1}^{\N_2}e^{i{k_2}\wedge p^{(2)}_j \eps^{(2)}_j}\right)
\left(\prod_{j=1}^{\N_3}e^{i{k_3}\wedge p^{(3)}_j \eps^{(3)}_j}
e^{ik_1\wedge p^{(3)}_j} \right),
\eea
where
\bea
\eps^{(1)}=0,1\ , \quad \eps^{(2)}=0,-1\ , \quad \eps^{(3)}=0,1\ . 
\nn
\eea

Inferring from Figs.~4 and 5, we extract the overall Filk phase-factor 
(which ought to show up in the $\star$-products of Eq.(\ref{GO})) as
a product of three copies ($a=1,2,3$) of  
\bea   
{\cal P}_{\rm t}{1\over \N_a!}\left
[\int_0^{\T_a}d\tau^{(a)}{\hat\Phi}(p)\right]^{\N_a}
&\ra& \sum_{\{\nu\}}\sum_{i_1<i_2<\cdots<i_{\N_a}} 
V^{(a)}_{i_1} \cdots V^{(a)}_{i_{\N_a}}\, \nn\\
&\times&\exp\left[\,{i\over4}\sum_{k<l}^{\N_a} p^{(a)}_{i_k}
\wedge p^{(a)}_{i_l} \left(\nu^{(a)}_{i_k}+\nu^{(a)}_{i_l}\right) 
\epsilon\left(\tau^{(a)}_{i_k}-\tau^{(a)}_{i_l}\right)\,\right]\ ,\label{ruleV}
\eea
multiplied by the moduli-independent phase-factor
\bea   
\label{xi}
\Xi\left(p^{(a)}_i \right) \!\! = 
\exp\left[\, -{i\over2} \sum_{a=1}^3 \left(\sum_{i,j=1}^{\N_a}
\left({1\over2}+\eta^{(a)}_i \right)
\left({1\over2}-\eta^{(a)}_j \right)p^{(a)}_i\wedge p^{(a)}_j 
+ {1\over3}\sum_i^{\N_a}\sum_j^{\N_{a+1}}p^{(a)}_i\wedge p^{(a+1)}_j
\right)\right].
\eea
Here, as shown in {}~Figs.~3(b) and 3(c), we have introduced
another set of insertion-specific sign factors:
\bea
\nu_j^{(a)} = (-)^{a+1} - 2\eps^{(a)}_j = - 2 \eta_j^{(a)}\ ~ . 
\eea
Note that (See Fig. 3(c))
\bea
\eta^{(1)}=\varepsilon^{(1)}-\frac{1}{2}\ ,\quad
\eta^{(2)}=\varepsilon^{(2)}+\frac{1}{2}\ ,\quad
\eta^{(3)}=\varepsilon^{(3)}-\frac{1}{2}\ 
\label{eeta}
\eea
defines the most symmetric assignment of the sign factor.
In Eq.(\ref{ruleV}), $\sum_{ \{\nu \} } $ denotes summation over all 
possible $2^{\N_a}$ choices of each $\nu_j^{(a)}$, corresponding two
inequivalent ways (distinguished by different Moyal's phase-factor)
of attaching $\Phi_0$-field on $a$-th internal propagator. 

Eqs.(\ref{ruleV}, \ref{xi}) are structurally similar to Filk's phase-factors
encountered in N-point Green functions at one-loop \cite{ours2}. As we
will elaborate in section 5, Eqs.(\ref{ruleV}, \ref{xi}) are intimately
tied with the generalized $\star$-product and, upon summing over N, 
re-expression of the two-loop 
effective action in terms of the scalar open Wilson lines.

\vspace{8mm}
\begin{minipage}[htb]{8.0cm} 
\begin{center}
\input{figw1.pictex}
\end{center}
{\bf Figure 4:} {\sl 
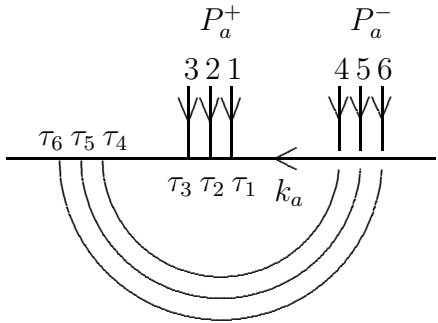
On an internal line, Filk's phase-factor for the planar external insertions 
has an opposite sign to the one for nonplanar external insertions.}
\end{minipage}
\hspace{10mm}
\vspace{8mm}
\begin{minipage}[htb]{6cm} 
\begin{center}
\input{figw2.pictex}
\end{center}
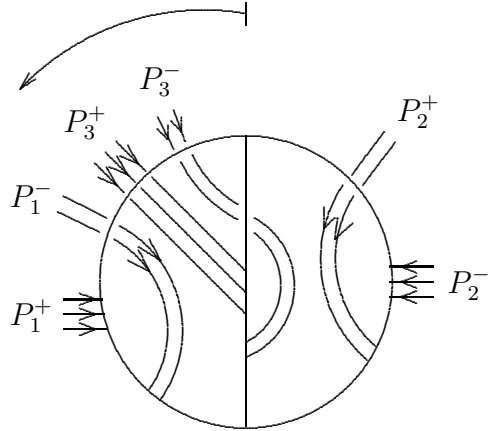
{\bf Figure 5:} {\sl Assignment of the overall Filk's phase-factor.}
\end{minipage}

The phase-factors Eqs.(\ref{ruleV},\ref{xi}) are accounted for as follows.
First, according to the phase-factor prescription adopted as above 
(see {}~Fig.~4), 
one ought to arrange crossing of nonplanar insertions (nonzero $\eps^{(a)}$) 
{\em prior to} any planar insertions. Moreover, lack of phase-factors for
cyclically ordered $\Phi_0$-field insertions implies that, graphically, 
these insertions do not cross one another, explaining the origin of the
phase-factor in Eq.(\ref{ruleV}) \cite{liu1,liu2, ours1, ours2}, 
which involves the 
ordering factor $\epsilon \left(\tau^{(a)}_{i_k}-\tau^{(a)}_{i_l} \right)$.
Next, as readily understood from Fig.~5, additional Filk's phase-factors 
among the partial sum of momenta $P^\pm_a$'s,
\bea
P^{\pm}_a = \sum_{j=1}^{\N_a}\left(\,{1\pm \nu_j^{(a)}\over2}\right)\,
p^{(a)}_j \qquad {\rm and} \qquad
P_a = P_a^{+}+P_a^{-} = \sum_{j=1}^{\N_a}p^{(a)}_j,
\nn
\eea
needs to be included. Again imposing the condition that insertions on each
boundary do not cross one another, ordering of $P^\pm_a$'s, and hence
the Filk's phase-factors are well-defined. For instance, in {}~Fig.~5, 
we have prescribed the {}~Filk ordering as   
$P^{-}_3$, $P^{+}_3$,  $P^{-}_1$, $P^{+}_1$, $P^{-}_2$, and $P^{+}_2$.
The resulting phase-factors are collected into
\bea
\label{xi0}
\Xi\left(p^{(a)}_i\right) \,\, = \,\, 
\exp\left[\, -{i\over2}\sum_{a=1}^3 \left(
(-1)^a P^{+}_a \wedge P^{-}_a +{1\over3}P_a \wedge P_{a+1}
\right) \right] ,
\eea
yielding, when expanded in terms of individual $p^{(a)}_i$'s, precisely 
the $\Xi$-factor, Eq.(\ref{xi}). Note that the $\Xi$ phase-factor
vanishes identically at one-loop. 

Putting together all the Filk phase-factors, the N-point Green function 
is expressible as:
\bea
\Gamma_\N = {1\over24}\sum_{\N=0}^\infty (-\lambda)^{2+\N}
\sum_{\N_1,\N_2,\N_3=0}^\N
\sum_{\sigma(\N_1,\N_2,\N_3)}
\sum_{\{\nu\}} C_{\{\N\}} \Gamma_{(\N_1,\N_2,\N_3)} \ ,
\nn
\eea
where $C_{\{\N\}}$ denotes the combinatoric factor associated with
marked N points on the two-loop graph, $\N_1 + \N_2 + \N_3 = \N$, and
\bea
\Gamma_{(\N_1,\N_2,\N_3)} &=&
\prod_{a=1}^3\int\limits_0^\infty \d\T_a \int\limits_0^{\T_a}
\prod_{i=1}^{\N_a}d\tau^{(a)}_i \, \Xi\left( p^{(a)}_i \right)
\exp\left[\,{i\over4}\sum_{k<l}^{\N_a} p^{(a)}_{k}
\wedge p^{(a)}_{l} \left(\nu^{(a)}_{k}+\nu^{(a)}_{l}\right) 
\epsilon\left(\tau^{(a)}_{k}-\tau^{(a)}_{l}\right)\,\right] \nn\\
&\times& \!\!\! \int \d^d x_1 \d^d x_2 
\int_{\scriptstyle y_a(0)=x_1 
        \atop\scriptstyle y_a(\T)=X_a}{\cal D}y_a\left(\tau^{(a)}\right)
\exp\left[\,i\sum_{j=1}^{\N_a}p^{(a)}_j y_a\left(\tau_j^{(a)}\right)\,\right] 
\label{nptgf} \\
&\times& \!\!\! \int{\cal D} k_a \left(\tau^{(a)}\right) 
\exp\left[ -\int^{\T_a}_0 \!\! (\,k^2_a+m^2 - ik_a\dot{y}_a \,)
\d \tau^{(a)}\right] \prod_{j=1}^{\N_a}{}^\prime
\exp\left[\,ik_a\left(\tau_j^{(a)}\right)
\wedge p^{(a)}_j\eps^{(a)}_j\right].
\nn
\eea
Here, for the planar part, we set $X_1 = X_2 = X_3 = x_2$, and, for 
nonplanar part, we take $(X_1,X_2,X_3)=(z,w,x_2)$ first and, after acting
the pseudo-differential operator $e^{ - i \partial_z \wedge \partial_w}$, 
set $z=w=x_2$.  
Because of the phase-factor convention adopted, the $k_a$-integration in
the last line of Eq.(\ref{nptgf}) take a slightly different form for $a=1$
and for $a=2,3$.  For $a=1$,
\bea
\int{\cal D}k_1\,\exp\left[-\int\limits_0^{\T_1} 
\left( k_1^2(\tau)-ik_1 \cdot {\dot y}_1(\tau) 
- ik_1(\tau)\wedge \left(
\sum_{j=1}^{\N_1} p^{(1)}_j \eps^{(1)}_j
+ \sum_{j=1}^{\N_3} p^{(3)}_j \right)
\delta \left(\tau-\tau^{(1)}_j\right)\right) \d\tau \right]\ ,
\nn
\eea
while, for $a=2,3$,  
\bea
\label{Ka}
\int{\cal D}k_a\,\exp\left[-\int\limits_0^{\T_a}\left(
k_a^2(\tau)-ik_a\cdot{\dot y}_a(\tau) 
- ik_a(\tau)\wedge \left( \sum_{j=1}^{\N_a}p^{(a)}_j 
\eps^{(a)}_j\right) 
\delta \left(\tau-\tau^{(a)}_j\right)\,\right) \d\tau\right]\ .
\eea
The notation $\prod^{\prime}$ in Eq.(\ref{nptgf}) is to emphasize the
extra contribution $\prod_{j=1}^{\N_3}e^{ik_1\wedge p^{(3)}_j}$ originating 
from the last phase-factor in Eq.\eq{addphase}.

After performing $k_a$ integrations, 
we obtain
\bea
\Gamma_{(\N_1,\N_2,\N_3)} &=&
\prod_{a=1}^3\int\limits_0^\infty \d\T_a \int\limits_0^{\T_a}
\prod_{i=1}^{\N_a}d\tau^{(a)}_i \Xi \left(p^{(a)}_i \right)
\exp\left[{i\over4}\sum_{k<l}^{\N_a} p^{(a)}_{k}
\wedge p^{(a)}_{l} \left(\nu^{(a)}_{k}+\nu^{(a)}_{l}\right) 
\epsilon\left(\tau^{(a)}_{k}-\tau^{(a)}_{l}\right)\right] \nn\\
&\times& \!\!\! \int \d^d x_1 \d^d x_2
\int_{\scriptstyle y_a(0)=x_1 
        \atop\scriptstyle y_a(\T)=X_a}{\cal D}y_a 
\exp\left[i\sum_{j=1}^{\N_a}p^{(a)}_j y_a\left(\tau_j^{(a)}\right)\right] 
\prod_{j=1}^{\N_a}{}^\prime \exp\left[
-{1\over2}{\dot y}_a(\tau_j^{(a)})\wedge p_j^{(a)}\eps^{(a)}_j
\right].
\nn
\eea
The functional integration over $y^a$ demonstrates manifestly that
the `noncommutative' scalar vertex operator ought to be identified with
\bea
V_j^{(a)}=\int_0^{\T_a} \d\tau 
\exp\left[\, i p_j \cdot \left(x^{(a)}(\tau) - 
{i\over2}\eps_j \wedge {\dot x}^{(a)}(\tau) \right) \,\right] ~ .
\nn
\eea
Performing the $y_a$ integral as in section 2, we finally obtain
\footnote{ The expression is arrangeable cyclic symmetrically if the
$\eta$'s defined in Eq.(\ref{eeta}) is used for the sign factors.}
\bea
\label{GON}
\Gamma_{(\N_1, \N_2, \N_3)} &=&
\prod_{a=1}^3\int\limits_0^\infty \d \T_a \int\limits_0^{\T_a}
\prod_{i=1}^{\N_a} \d\tau^{(a)}_i \Xi\left(p^{(a)}_i \right)
\exp\left[\,{i\over4}\sum_{k<l}^{\N_a} p^{(a)}_{k}
\wedge p^{(a)}_{l} \left(\nu^{(a)}_{k}+\nu^{(a)}_{l} \right) 
\epsilon \left(\tau^{(a)}_{k}-\tau^{(a)}_{l} \right)\right] \nn\\
&\times& \int \d^d x_1 \d^d x_2
\left(1\over4\pi \T_a\right)^{d \over 2}
\exp\left[\,{-(X_a-x_1)^2\over4 \T_a}\,\right]
\exp\left[\,i\sum_{j=1}^{\N_a}p^{(a)}_j x_1 \,\right] \nn\\
&\times&\exp\left[\,i(X_a-x_1) 
\sum_{j=1}^{\N_a}{}^{\hskip-1pt\prime}\,{1\over \T_a}\left(\,
\tau^{(a)}_j \cdot p^{(a)}_j+{i\over2}\wedge p^{(a)}_j 
\eps^{(a)}_j\,\right)\,\right]  \nn\\
&\times&\exp\Biggl[\,\,   {1\over2}\sum_{i,j=1}^{\N_a} 
p^{(a)}_i\cdot p^{(a)}_j\,
\left(\,\left|\tau^{(a)}_i-\tau^{(a)}_j \right|
-\left(\tau^{(a)}_i+\tau^{(a)}_j\right)
+ {2\tau^{(a)}_i\tau^{(a)}_j\over \T_a}\,\right) \nn\\
&& \hskip0.6cm -{i\over2}\sum_{i,j=1}^{\N_a}{}^{\hskip-4pt\prime}\, 
p^{(a)}_i\wedge p^{(a)}_j\,
\left( \,\eps^{(a)}_j\left(\,1-{\tau^{(a)}_i-\tau^{(a)}_j\over \T_a}\,\right)
+\eps^{(a)}_j\left(\,1-{\tau^{(a)}_i+\tau^{(a)}_j\over \T_a}
\,\right)\,\right)\nn\\
&& \hskip0.6cm -{1\over4 \T_a}\sum_{i,j=1}^{\N_a}{}^{\hskip-4pt\prime}\, 
p^{(a)}_i\circ p^{(a)}_j\,
\eps^{(a)}_i\eps^{(a)}_j\,\,\Biggr]\ .
\eea
It now remains to perform the $x_1, x_2$ integrals in
Eqs.(\ref{GP}, \ref{GNP}). In the next subsections, we will do so
for the planar and the nonplanar cases separately.
\subsection{Planar Part}
For the planar contribution, we set $z=w=x_2$, and perform the $x_1, x_2$ 
integrations \footnote{Details are sketched in Appendix A.}. 
We obtain the result as
\bea
\label{GPN}
\Gamma^{\bf P}_\N = {1 \over 24} (-\lambda)^{2+\N}\sum_{\N_1,\N_2,\N_3=0}^\N
\sum_\sigma \sum_{\{\nu\}}
(2\pi)^d \delta \left( \sum_{a=1}^3\sum_{i=1}^{\N_a}p_i^{(a)} \right)
\,C^{\bf P}_{\{\N\}} \, \Gamma^{\bf P}_{(\N_1,\N_2,\N_3)} 
\eea
with
\bea
\label{master1}
\Gamma^P_{(\N_1,\N_2,\N_3)}&=& {1\over(4\pi)^d}
\int\limits_0^\infty \prod_{a=1}^3 \d\T_a e^{-m^2 \T_a}
\Delta^{d \over2}(\T) \prod_{a=1}^3 \int\limits_0^{\T_a}
\prod_{i=1}^{\N_a} \d\tau^{(a)}_i \Xi\left( p^{(a)}_i \right)  \nn \\
&\times& 
\exp\left[\,{i\over4}\sum_{k<l}^{\N_a} p^{(a)}_{k}
\wedge p^{(a)}_{l} \left(\nu^{(a)}_{k}+\nu^{(a)}_{l} \right) 
\epsilon \left(\tau^{(a)}_{k}-\tau^{(a)}_{l} \right)\,\right] \nn\\
&\times& \exp\Biggl[\,{1\over2}\sum_{a=1}^3 
\sum_{i,j}^{\N_a} p^{(a)}_i \cdot 
G^{\bf P}_{aa}\left(\tau^{(a)}_i,\tau^{(a)}_j; \eta^{(a)}_i,\eta^{(a)}_j
\right) p_j^{(a)} \nn\\
&+&\sum_{a=1}^3\sum_i^{\N_a}\sum_j^{\N_{a+1}} p^{(a)}_i
G^{\bf P}_{aa+1} \left(\tau^{(a)}_i,\tau^{(a+1)}_j;\eta^{(a)}_i,\eta^{(a+1)}_j
\right) p^{(a+1)}_j \,\Biggr]\ .  
\eea
Here, the worldline correlators for the planar vacuum diagram are given by
\bea
G^{{\bf P}^{mn}}_{aa}
\left(\tau^{(a)}_i,\tau^{(a)}_j; \eta^{(a)}_i,\eta^{(a)}_j \right)  
&=& g^{mn}G^{\rm sym}_{aa}(\tau^{(a)}_i,\tau^{(a)}_j) 
-2i\Delta (\T) \theta^{mn}(\T_{a+1}+\T_{a+2})\tau^{(a)}_i \eta^{(a)}_j\nn\\
&&\quad +\frac{1}{4} \Delta(\T) (-\theta^2)^{mn}(\T_{a+1} + \T_{a+2})
   \left(\eta_{i}^{(a)} - \eta_{j}^{(a)}\right)^{2}\ ,
\nn
\eea
and 
\bea
G^{{\bf P}^{mn}}_{aa+1} 
\left(\tau^{(a)}_i,\tau^{(a+1)}_j; \eta^{(a)}_i,\eta^{(a+1)}_j \right)  
&=& g^{mn}G^{\rm sym}_{aa+1} \left(\tau^{(a)}_i,\tau^{(a+1)}_j \right) \nn\\
&& \hskip-4.5cm +\,\,  i\Delta(\T) \theta^{mn} \left[\, \T_{a+2}
  \left(\tau_{i}^{(a)}\eta_{j}^{(a+1)} -\eta_{i}^{(a)}\tau_{j}^{(a+1)}\right) 
   + \frac{1}{2} \T_{a+1} \tau_{i}^{(a)}+\frac{1}{2} \T_{a} \tau_{j}^{(a+1)}
   \,\right] \nn \\
&&\hskip-4.5cm +\,\,  \frac{1}{4} \Delta(\T) (-\theta^2)^{mn}
    \left[\, \left(\eta_{i}^{(a)}+\frac{1}{2} \right)^2 \T_{a+1} 
    + \left(\eta_{j}^{(a+1)}-\frac{1}{2} \right)^2 \T_{a} 
+\left(\eta_{i}^{(a)} + \eta_{j}^{(a+1)}\right)^2 \T_{a+2} \,\right] \ .
\nonumber
\eea
In this expression, as explained in Appendix A, we have reversed ordering 
of the $\tau$-variable via $\tau^{(a)} \rightarrow \T_a - \tau^{(a)}$.
Consequently, the sign of the $\epsilon\left( \tau^{(a)}_k - \tau^{(a)}_\ell 
\right)$-dependent phase in Eq.(\ref{master1}) is reversed  compared to, 
for example, that in Eq.(\ref{GPN}). It is to be noted that the last term,
proportional to $(-\theta^2)^{mn}$, is structurally the same as the 
$G^{\rm sym}_{a a+1}$ worldline correlator, Eq.(\ref{Gs12}), except that
the $\Phi_0$-insertion dependence is through the sign-factors (instead of
the insertion moduli parameters). 
\subsection{Nonplanar Part}
In this case, we need to insert the pseudo-differential operator
$e^{- i \partial_z \wedge \partial_w}$ first, and then perform the $x_1, x_2$ 
integrations. 
The operator insertion amounts effectively to the $\star$-product among 
the background-dependent propagators. See Appendix A for details. 
The insertion also causes the $\Delta(\T_a)$  factor being replaced by 
$\Delta_\theta(\T_a)$: 
\bea
 \Delta_\theta (\T_a) = 
\left(\T_1 \T_2+\T_2 \T_3+ \T_3 \T_1 -\frac{\theta^2}{4} \right)^{-1}\ .
\nn
\eea
The final result is expressible as follows:
\bea
\label{GNPN}
\Gamma^{\bf NP}_\N = {1 \over 24} (-\lambda)^{2+\N}
\sum_{\N_1,\N_2, \N_3=0}^\N \sum_\sigma \sum_{\{\nu\}}
(2\pi)^d \delta \left(\sum_{a=1}^3\sum_{n=1}^{\N_a}p_n^{(a)} \right)
\,C^{\bf NP}_{\{\N\}} \, \Gamma^{\bf NP}_{(\N_1,\N_2,\N_3)} 
\eea
with
\bea
\label{master2}
\Gamma^{\bf NP}_{(\N_1,\N_2,\N_3)}&=& {1\over(4\pi)^d}
\int\limits_0^\infty \prod_{a=1}^3 \d \T_a e^{-m^2 \T_a}
(\Delta_\theta)^{d\over2}(\T_a) 
\prod_{a=1}^3 \int\limits_0^{\T_a}
\prod_{i=1}^{\N_a}d\tau^{(a)}_i\Xi\left(p^{(a)}_i \right) \nn \\
&\times&
\exp\left[\,{i\over4}\sum_{k<l}^{\N_a} p^{(a)}_{k}
\wedge p^{(a)}_{l} \left(\nu^{(a)}_{k}+\nu^{(a)}_{l} \right) 
\epsilon \left(\tau^{(a)}_{k}-\tau^{(a)}_{l}\right)\right] \nn\\
&\times& \exp\Biggl[\, {1\over2}\sum_{a=1}^3 \sum_{i,j}^{\N_a} 
p^{(a)}_i G^{{\bf NP}}_{aa} 
\left(\tau^{(a)}_i,\tau^{(a)}_j; \eta^{(a)}_i,\eta^{(a)}_j\right)
p^{(a)}_j \nn\\
&& +\sum_{a=1}^3\sum_i^{\N_a}\sum_j^{\N_{a+1}} 
p^{(a)}_i G^{{\bf NP}}_{aa+1} 
\left(\tau^{(a)}_i,\tau^{(a+1)}_j;\eta^{(a)}_i,\eta^{(a+1)}_j\right) 
p^{(a+1)\nu}_j
\,\Biggr]\ .  
\eea
Here, the worldline correlators for the nonplanar two-loop vacuum 
diagram are defined as:
\bea
\label{npfinala}
G^{{\bf NP}^{mn}}_{aa}
\left(\tau^{(a)}_i,\tau^{(a)}_j; \eta^{(a)}_i,\eta^{(a)}_j\right)  
&=& g^{mn}G^{\rm sym}_{\theta\, aa} \left(\tau^{(a)}_i,\tau^{(a)}_j \right) 
-2i\Delta_\theta(\T) \theta^{mn}(\T_{a+1}+\T_{a+2})\tau^{(a)}_i \eta^{(a)}_j \nn\\
&&\hskip-4.5cm 
+{1 \over 4} \Delta_\theta(\T)(-\theta^2)^{mn}(\T_{a+1} + \T_{a+2})
   \left(\eta_{i}^{(a)} - \eta_{j}^{(a)}\right)^{2} 
-{i \over 4} \Delta_\theta(\T) (\theta^3)^{mn}
\left(\eta^{(a)}_i-\eta^{(a)}_j\right),
\eea
and 
\bea
\label{npfinalb}
G^{{\bf NP}^{mn}}_{aa+1}
\left(\tau^{(a)}_i,\tau^{(a+1)}_j; \eta^{(a)}_i,\eta^{(a+1)}_j\right)  
&=& g^{mn}G^{\rm sym}_{\theta\, aa+1} \left(\tau^{(a)}_i,\tau^{(a+1)}_j\right) 
\nn\\
&&\hskip-140pt +\,\,i \Delta_\theta(\T) \theta^{mn} 
  \left[\, \T_{a+2}\left(\tau_{i}^{(a)} \eta_{j}^{(a+1)}
   - \eta_{i}^{(a)} \tau_{j}^{(a+1)}\right)-\frac{1}{2} \T_{a+1} 
   \tau_{i}^{(a)}-\frac{1}{2} \T_{a}  \tau_{j}^{(a+1)}
   +\tau_{i}^{(a)}\tau_{j}^{(a+1)} \right] \nn\\
&&\hskip-140pt+{1 \over 4} \Delta_\theta(\T) (-\theta^2)^{mn}
    \left[\, \left(\eta_{i}^{(a)}-\frac{1}{2}\right)^2 \T_{a+1} 
    + \left(\eta_{j}^{(a+1)}+\frac{1}{2}\right)^2 \T_{a}
   +\left(\eta_{i}^{(a)} + \eta_{j}^{(a+1)}\right)^2 \T_{a+2} \right] \nn\\
&&\hskip-140pt+{1 \over 2} \Delta_\theta(\T) (-\theta^2)^{mn}
   \left[\,\left(\eta_{i}^{(a)}-\frac{1}{2}\right) \tau_{j}^{(a+1)}
   - \left(\eta_{j}^{(a+1)}+\frac{1}{2}\right) \tau_{i}^{(a)} \right]\nn\\
&&\hskip-140pt+{i\over4} \Delta_\theta(\T) (+\theta^{3})^{mn} 
   \left[\,\left(\eta_{i}^{(a)}+\frac{1}{2}\right)
   \left(\eta_{j}^{(a+1)}-\frac{1}{2} \right)
    +\frac{1}{3}\right]
+\frac{i}{3} \theta^{mn} \ .
\eea
Moreover, $G^{\rm sym}_{\theta\, ab}$ are defined by 
replacing $\Delta(\T)$ by $\Delta_\theta(\T)$ in $G^{\rm sym}_{ab}$. 

A notable difference of the nonplanar case, as compared to the planar case, 
is that the worldline propagators comprise of ${\cal O}(\theta^3)$ 
contribution, and the contribution is independent of the $\Phi_0$-field
insertion moduli. In Appendix B, we illustrate the contribution explicitly
by computing several two-loop nonplanar Feynman diagrams.   
\section{Generalized Star Products and Open Wilson Lines}
\setcounter{section}{5}
\setcounter{equation}{0}
\indent 
In the last section, we have computed systematically the two-loop
contribution to the N-point, one-particle-irreducible Green functions
in noncommutative $\lambda [\Phi^3]_\star$-theory. We have adopted the
worldline formulation for the computation in order to make contact with
the Seiberg-Witten limit from the underlying string theories. 

According to the conjecture \cite{sjrey}, once resummed over N with 
appropriate combinatoric factors, the resulting two-loop effective action 
ought to be expressible in terms of open Wilson lines, viz. the effective
action represents interactions among the noncommutative dipoles. 
A prerequisite to emergence of such structures is that the effective
action ought to be a functional defined in terms of generalized 
$\star$-products of the elementary field $\Phi$, much as in the one-loop
effective action. There, the generalized $\star$-product was expressible
in momentum-space representation in terms of the following kernel:
\bea
J_\N := \int_0^1 \cdots \int_0^1 \d \tau_1 \cdots \d \tau_\N
\exp \left[ - i \sum_{i=1}^\N \tau_i p_i \wedge k
+{i \over 2} \sum_{i<j=1}^\N \epsilon(\tau_i - \tau_j)
p_i \wedge p_j \right].
\label{oneloopgenstar}
\eea
In this section, for
both the planar and the nonplanar contributions, we will indicate
briefly how the generalized $\star$-products come about and in what
sense they are resummable into open Wilson lines. 

Begin with the the planar contribution, Eq.(\ref{GPN}), corresponding
in underlying string theory to the worldsheet of genus zero with 
three holes. Were if it reexpressible as representing cubic interaction
among dipoles, each boundary of the hole ought to be identified with an 
open Wilson line whose size in spacetime is proportional to the total 
momentum inserted along the boundary. Thus, introduce the $a$-th
boundary momentum as:
\bea
 k_a = \sum_{i=1}^{\N_a} \left( \frac{1}{2} - \eta_i^{(a)} \right) p_i^{(a)}
    + \sum_{i=1}^{\N_{a+1}} \left( \frac{1}{2} + \eta_i^{(a+1)} \right)
 p_i^{(a+1) } ~ ,
\qquad (a=1,2,3).
\nn
\eea
They constitute a direct two-loop counterpart of the momentum $k$ in 
Eq.(\ref{oneloopgenstar}).  
Remarkably, we have found that the second exponential in  Eq.(\ref{master1}) 
is simplified into a suggestive form:
\bea
&& \exp \left[ 
   \Delta(\T) \sum_{a=1}^3 \left\{ \T_{a+2} k_a \wedge
  \left( \sum_{i=1}^{\N_{a+1}} \tau_i^{(a+1)} p_i^{(a+1)}
 - \sum_{i=1}^{\N_a} \tau_i^{(a)} p_i^{(a)} \right) 
  - \frac{1}{4} \T_{a+2} k_a \circ k_a \right\} \right] 
\nn \\
&\rightarrow& 
\prod_{a=1}^3 \exp \left[ - i \Delta(\T)
\sum_{i=1}^{\N_a} \tau^{(a)}_i p^{(a)}_i \wedge
\left( \T_{a+2} k_a - \T_a k_{a+1} \right) \right]
\exp \left[ - {1 \over 4} \Delta(\T) \T_{a+2} k_a \circ k_a \right],
\label{exponential}
\eea
where we have first taken the low-energy and 
large noncommutativity limit, considered already in \cite{ours1}. 
In the last expression, the phase-factor of the first exponential 
is readily seen as the counterpart of the first term in 
Eq.(\ref{oneloopgenstar}). Thus, combining the first exponential of
Eq.(\ref{exponential})
with the $\Xi$-factor and the first exponential in Eq.(\ref{master1}),
we readily find that kernels reminiscent of Eq.(\ref{oneloopgenstar})
ought to follow. On the other hand, the second exponential of
Eq.(\ref{exponential}) is quadratic in $\theta^{mn} k_n$, and hence
act as the Gaussian damping factor turning, for each $a$-th boundary,
UV dynamics into IR dynamics.   
Results including these aspects will be reported in a separate paper elsewhere
\cite{pants}.

The nonplanar contribution, Eq.(\ref{GNPN}), also exposes new features.
The contribution corresponds to, in underlying string theory, the worldsheet 
of genus one with a hole, yielding homology cycles with nontrivial 
intersections.  As such, compared to the planar contribution Eq.(\ref{GPN})
(for which there is no nontrivial homology cycle), all extra terms present
in Eq.(\ref{GNPN}) would be attributable to the existence of the intersecting 
homology cycles \cite{chu}. An immediate question is whether the
open Wilson lines play an important role --- for example, whether the nonplanar
contribution, Eq.(\ref{GNPN}), is also expressible entirely in terms of the
open Wilson lines. If so, these homology cycles ought to describe `worldsheet' 
of  {\sl virtual} open Wilson lines, and hence would disclose {\sl quantum 
aspects} of the noncommutative dipoles in noncommutative field theories.  
We are currently investigating various aspects including these issues and
will report them elsewhere. 

\section*{Acknowledgements}
\indent 
We thank to Costas Bachas, Sangmin Lee, Peter Mayr, Ashoke Sen, and Jung-Tay 
Yee for enlightening discussions. SJR acknowledges warm hospitality of 
Spinoza Institute at Utrecht 
University, Department of Mathematics \& Physics at Amsterdam University,
Summer Institute at Yamanashi-Japan, \'Ecole de Physique -- Les Houches,
and Theory Division at CERN, where various parts of this project were 
accomplished.
\newpage
\appendix
\section*{Appendix}
\section{Derivation of Eq.(\ref{master1}) and Eq.(\ref{master2})}
\label{ap1}
\indent
As a starting point, we rewrite Eq.\eq{GON} and apply the change of 
variable $Y=X_a -x_1$ for the $x_2$ integration (the $z$ and $w$ 
dependences are retained for the nonplanar case). In what follows, 
we will omit the overall energy-momentum $\delta$-function arising from 
\bea
\prod_{a=1}^3\prod_{j=1}^{\N_a} 
\int \d^d x_1  e^{ip^{(a)}_j x_1}=
(2\pi)^d \delta\left(\,\sum_{a=1}^3\sum_{j=1}^{\N_a}p^{(a)}_j\,\right) \ .
\nn
\eea
Introduce functions $f_a$; $a=1,2,3$:
\bea
f_a(Y)=\exp \left[\, -{1\over4\T_a}Y^2 + i Y \cdot  q_a \, \right] 
\quad {\rm where} \quad  
q_a^m = \sum_{j=1}^{\N_a}{}^{\hskip-1pt\prime}\,{1\over \T_a}\left(\,
\tau^{(a)}_j p^{(a)m}_j+{i\over2}\theta^{mn}p^{(a)n}_j 
\eps^{(a)}_j\,\right) \ .
\nn
\eea
Note that, for notational simplicity, we have suppressed dependence of
$f_a$'s on the insertion moduli, $\tau^{(a)}_j$'s. 
Introduce also functions, $H_0$ and $H_1$:
\bea
H_0 = {i\over4}\sum_{a=1}^3\sum_{k<l}^{\N_a} p^{(a)}_{k}
\wedge p^{(a)}_{l}\left(\nu^{(a)}_{k}+\nu^{(a)}_{l}\right) 
\epsilon\left(\tau^{(a)}_{k}-\tau^{(a)}_{l} \right) \ ,
\nn
\eea
and
\bea
H_1 &=& {1\over2}\sum_{a=1}^3\sum_{i,j=1}^{\N_a} 
p^{(a)}_i\cdot p^{(a)}_j\,\left(\,
\left\vert\tau^{(a)}_i-\tau^{(a)}_j\right\vert
- \left( \tau^{(a)}_i+ \tau^{(a)}_j \right)
+ {2\tau^{(a)}_i\tau^{(a)}_j\over \T_a}\,\right)\nn\\
&-&{i\over2}\sum_{a=1}^3\sum_{i,j=1}^{\N_a}{}^{\hskip-4pt\prime}\, 
p^{(a)}_i\wedge p^{(a)}_j\,
\left(\,\eps^{(a)}_j\left[1-{\tau^{(a)}_i-\tau^{(a)}_j\over \T_a}\right]
+\eps^{(a)}_j\left[1-{\tau^{(a)}_i+\tau^{(a)}_j\over \T_a} \right]
\,\right) \nn\\
&-&\sum_{a=1}^3\sum_{i,j=1}^{\N_a}{}^{\hskip-4pt\prime}\,{1\over 4 \T_a} 
p^{(a)}_i\circ p^{(a)}_j\,
\eps^{(a)}_i\eps^{(a)}_j  \ .
\nonumber
\eea
We were then able to express the $\N$-point Green functions defined in 
Eqs.(\ref{GPN}, \ref{GNPN}) as  
\bea
\label{gpn}
\Gamma^{\bf P}_{(\N_1,\N_2,\N_3)} = 
\prod_{a=1}^3\int\limits_0^\infty \d\T_a 
\left({1\over4\pi \T_a}\right)^{d\over2} e^{-m^2 \T_a}
\int\limits_0^{\T_a}
\prod_{i=1}^{\N_a}\d \tau^{(a)}_i e^{H_0+H_1}
\int \d^d Y f_1(Y)f_2(Y)f_3(Y) \ ,
\eea
and
\bea
\label{gnpn}
\Gamma^{\bf NP}_{(\N_1,\N_2,\N_3)} =
\prod_{a=1}^3\int\limits_0^\infty \d \T_a 
\left({1\over4\pi \T_a}\right)^{d \over2} e^{-m^2 \T_a}
\int\limits_0^{\T_a}
\prod_{i=1}^{\N_a}d\tau^{(a)}_i e^{H_0+H_1}
\int \d^d Y 
\left( f_1(Y) \star f_2(Y)\right) f_3(Y).
\eea
\subsection{Planar Case Eq.(\ref{master1})}
{}~For the planar case, Eq.\eq{gpn}, integration over $Y^m$ is expressible
as:
\bea
\Gamma^{\rm n}_{(\N_1,\N_2,\N_3)} = \prod_{a=1}^3 \int\limits_0^\infty  \d\T_a 
\left({1\over4\pi \T_a}\right)^{d \over2} e^{-m^2\T_a}
\int\limits_0^{\T_a}
\prod_{i=1}^{\N_a}\d \tau^{(a)}_i \, e^{H_0} e^{H_1+H_2}\ ,
\nonumber
\eea
where
\bea
H_2 = -\Delta(\T) \sum_{a=1}^3 \sum_{b=1}^3 
\sum_{j=1}^{\N_a}{}^{\hskip-2pt\prime}\,
\sum_{k=1}^{\N_b}{}^{\hskip-2pt\prime}\,
{\T_1\T_2\T_3\over \T_a\T_b}
\left( p^{(a)}_j\cdot p^{(b)}_k\tau^{(a)}_j\tau^{(b)}_k+
       p^{(a)}_j\wedge p^{(b)}_k\tau^{(a)}_j\eps^{(b)}_k
       -{1\over4} p^{(a)}_j\circ p^{(b)}_k\eps^{(a)}_j\eps^{(b)}_k
\right) .
\nonumber
\eea
The $e^{H_0}$ part reproduces the first exponential in Eq.(\ref{master1}).
The exponent of $e^{H_1 + H_2}$ is expandable explicitly as:
\bea\label{plan}
(H_1+H_2) &= & \frac{1}{2}\sum_{a=1}^{3}\sum_{i,j=1}^{\N_{a}}
  \Biggl\{p_{i}^{(a)}\cdot p_{j}^{(a)}
G^{\rm sym}_{aa}(\tau^{(a)}_i,\tau^{(a)}_j) 
- 2i p_{i}^{(a)}\wedge p_{j}^{(a)}\eps_{j}^{(a)} \nn\\
&&\hspace{13mm} +2i \Delta (\T) p_{i}^{(a)}\wedge p_{j}^{(a)}
  (\T_{a+1} + \T_{a+2}) \tau_{i}^{(a)}\eps_{j}^{(a)} \nn\\
&&\hspace{13mm} +{1 \over 4} \Delta(\T) p_{i}^{(a)}\circ p_{j}^{(a)}
(\T_{a+1} + \T_{a+2}) 
\left(\eps_{i}^{(a)} - \eps_{j}^{(a)}\right)^{2} \Biggl\}\nn\\
&&+\sum_{a=1}^{3}\sum_{i=1}^{\N_{a}}\sum_{j=1}^{\N_{a+1}}
   \Biggl\{ p_i^{(a)}p_j^{(a+1)} 
G^{\rm sym}_{aa+1} \left(\tau^{(a)}_i,\tau^{(a+1)}_j\right) \nn\\
&& \hspace{13mm}- i \Delta(\T) p_{i}^{(a)}\wedge p_{j}^{(a+1)}
   \left(\tau_{i}^{(a)}\eps_{j}^{(a+1)}
    - \eps_{i}^{(a)}\tau_{j}^{(a+1)}\right) \T_{a+2} \nn\\
&&\hspace{13mm} - {1 \over 4} \Delta (\T) 
p_{i}^{(a)}\circ p_{j}^{(a+1)}
\left[(\eps_{i}^{(a)})^2 \T_{a+1} + (\eps_{j}^{(a+1)})^2 \T_{a}
+(\eps_{i}^{(a)} + \eps_{j}^{(a+1)})^2 \T_{a+2}\right]\Biggl\}\nn\\
&&-i\Delta(\T) \sum_{i=1}^{\N_3}\sum_{j=1}^{\N_1}p_{i}^{(3)}\wedge p_{j}^{(1)}
    \left[(\T_2+\T_3)\tau_{j}^{(1)} -\T_2\tau_{i}^{(3)}\right]\nn\\
&&-i\Delta(\T)\sum_{i=1}^{\N_2}\sum_{j=1}^{\N_3} p_{i}^{(2)}\wedge p_{j}^{(3)}
     \left(\T_3\tau_{i}^{(2)} + \T_{2}\tau_{j}^{(3)}\right) 
   -i\sum_{i=1}^{\N_1}\sum_{j=1}^{\N_3}p_{i}^{(1)}\wedge p_{j}^{(3)}\nn\\
&& +\frac{1}{2}\Delta(\T) \sum_{i=1}^{\N_1}\sum_{j=1}^{\N_3}
   p_{i}^{(1)}\circ p_{j}^{(3)}\left[ 
    -(\T_2+\T_3)\eps_{i}^{(1)}-\T_2\eps_{j}^{(3)}+
     \frac{1}{2}(\T_2+\T_3) \right]\nn\\
&&+\frac{1}{2}\Delta(\T) \sum_{i=1}^{\N_2}\sum_{j=1}^{\N_3}
  p_{i}^{(2)}\circ p_{j}^{(3)}\left[\T_3\eps_{i}^{(2)}- 
  \T_2\eps_{j}^{(3)}+\frac{1}{2}(\T_2+\T_3) \right].
\eea
Here, the $G^{\rm sym}_{ab}$ propagators are precisely those defined in 
Eqs.(\ref{Gsaa}, \ref{Gs12}) 
and, at appropriate places, we have applied the following energy-momentum 
conservation identity:
\begin{eqnarray}
 & \sum_{a=1}^3\sum_{i,j=1}^{\N_a}p^{(a)}_i\theta^M p^{(a)}_j 
C_a \left(\tau^{(a)}_i\right)^N
  =   \nonumber \\
 & \sum_{a=1}^3\sum_{i=1}^{\N_a}\sum_{j=1}^{\N_{a+1}}
p^{(a)}_i\theta^M p^{(a+1)}_j \left(\, -C_a (\tau^{(a)}_i)^N 
+(-1)^{M+1} C_{a+1} \left(\tau^{(a+1)}_j\right)^N \,\right)\ , 
\label{moform} 
\end{eqnarray}
where $M$ and $N$ are integer-valued, and the $C_a$$(a=1,2,3)$ are 
$\tau$-independent constants. The second exponential in Eq.(\ref{master1})
is then obtained, after straightforward and tedious algebra, from reversing 
the insertion moduli as $\tau^{(a)}\ra \T_a - \tau^{(a)}$, applying the 
formula, Eq.\eq{moform}, and rearranging them into cyclically symmetric form. 
Finally, the factor $\Xi$ in Eq.(\ref{master1}) follows from moduli-independent
phase-factors in Eq.(\ref{plan}).  
\subsection{Nonplanar Case Eq.(\ref{master2})}
For the nonplanar case Eq.\eq{gnpn}, we'll need first to act the
pseudo-differential operator $e^{-i\der_z\wedge\der_w}$ acting on $f_a$'s: 
\bea
W\equiv f_1(Y) \star f_2(Y)
= \left. f_1(z^m - i\theta^{mn}\der_w^n)f_2(w)\right\vert_{z=w=Y}\ .
\nn
\eea
We have computed this by taking the Fourier integral representation:
\bea
f_2(w)=\left({\T_2\over\pi}\right)^{d\over2}
\int\limits_{-\infty}^\infty \d^d k \,
e^{-\T_2(k+q_2)^2}e^{-ik\cdot w}.
\nn
\eea
The result turns out expressible as
\bea
W = \left(\T_2\right)^{d \over2}(\mbox{det}A)^{-1/2}
e^{{1\over4}J^{\rm T}A^{-1}J}
e^{-\T_2 q_2^2}f_1(Y) \ ,
\nn
\eea
where $A$ and $J$ refer to $(d \times d)$ matrix and $d$-component vector,
respectively:
\bea\label{matA}
A^{mn} = \T_2 \delta^{mn}- {1\over 4 \T_1}(\theta^2)^{mn}\ ,
\nn
\eea
\bea\label{matC}
J^m = \Bigl(-2\T_2q_2^m + i\theta^{mn} q_1^n\Bigr) + 
\left({-\theta\over2\T_1} - i\right)^{mn} Y^n 
\equiv j^m + C^{mn}Y^n \ .
\label{mats}
\eea
Inserting these expressions into Eq.\eq{gnpn}, and then integrating
over $Y$'s, we obtain 
\bea
\label{gnpnnew}
\Gamma^{NP}_{(\N_1,\N_2,\N_3)} &=& \prod_{a=1}^3\int\limits_0^\infty \d\T_a 
\left({1\over4\pi \T_a}\right)^{d \over2} e^{-m^2 \T_a} \int\limits_0^{\T_a}
\prod_{i=1}^{\N_a} \d\tau^{(a)}_i \, e^{H_0} e^{H_1+H'_2}\nn\\
&& \hskip2cm (\pi \T_2)^{d\over2}(\mbox{det}A)^{-1/2}
(\mbox{det}{\tilde A})^{-1/2} \ ,
\eea
where the new exponent is
\bea
H'_2 ={1\over4}j^{\rm T}A^{-1}j 
+ {\tilde J}^{\rm T}{\tilde A}^{-1}{\tilde J} - \T_2 q_2^2,
\nn
\eea
and $\widetilde{A}$ and $\widetilde{J}$ refer to newly combined 
$(d \times d)$ matrix and $d$-component vector, respectively: 
\bea
\tilde{A}&=& A_0  -{1\over4}C^{\rm T}A^{-1}C \qquad {\rm where} \qquad 
A_0={1\over4}(\T_1^{-1}+\T_3^{-1}) {\bf I} \ , \\
\tilde{J}^m&=& {1\over4}(C^{\rm T}A^{-1}j)^m+{i\over2}(q_1 + q_3)^m\ .
\nn
\eea
Using the relation
\bea
A \tilde{A}=A_0A-{1\over4}CC^{\rm T} = (4 \T_1 \T_3\Delta_\theta)^{-1}{\bf I}
\nn
\eea
and the fact that the matrices $A$, $A_0$ and $C$ commute one another, 
we have derived the following algebraic identities:
\bea
&&A^{-1}\Bigl(1+\T_1 \T_3 C\Delta_\theta (\T) C^{\rm T}\Bigr) 
= (\T_1+\T_3)\Delta_\theta (\T) \ , \label{form1} \nn \\
&&\tilde{A}^{-1}=4 \T_1 \T_2 \T_3\Delta_\theta (\T)
\left(1-{\theta^2\over4 \T_1 \T_2}\right)\ , \label{form2}\\
&&(\mbox{det}A)(\mbox{det}{\tilde A}) 
= (4 \T_1 \T_3\Delta_\theta)^{-d}\ . \label{form3}
\eea
In fact, due to Eq.\eq{form3}, the weight factor in Eq.\eq{gnpnnew} is 
proportional to $({Delta_\theta}(\T))^{d \over2}$, viz. 
\bea
\left(\prod_{a=1}^3{1\over4\pi \T_a}\right)^{d \over2} 
(\pi \T_2)^{d \over2}(\mbox{det}A)^{-1/2}
(\mbox{det}{\tilde A})^{-1/2} ={1\over(4\pi)^d } 
(\Delta_\theta(\T))^{d \over2} .
\nn
\eea
Using Eqs.\eq{form1}, \eq{form2} and \eq{moform}, 
the exponent $(H_1+H'_2)$ of the second exponential is expandable
explicitly as follows:  
\bea
\label{nnonp}
(H_1+H'_2)
&=&\frac{1}{2}\sum_{a=1}^{3}\sum_{i,j=1}^{\N_{a}}
  \Biggl\{ p_{i}^{(a)} \cdot p_{j}^{(a)}\Bigl[ 
\left\vert\tau_{i}^{(a)}-\tau_{j}^{(a)} \right\vert
-\Delta_\theta(\T) (\T_{a+1}+\T_{a+2})
\left(\tau_{i}^{(a)} - \tau_{j}^{(a)}\right)^{2} \Bigr]\nn\\
&& \hspace{5mm} -\frac{1}{4}  p_{i}^{(a)} \theta^2 p_{j}^{(a)}
\Delta_\theta(\T) (\T_{a+1}+ \T_{a+2})
    \left(\eps_{i}^{(a)}- \eps_{j}^{(a)}\right)^{2}\Biggr\}\nn\\
&&+\sum_{a=1}^{3}\sum_{i=1}^{\N_{a}}\sum_{j=1}^{\N_{a+1}} \biggl\{
    p_{i}^{(a)} \cdot p_{j}^{(a+1)}\Biggl[
\left(\tau_{i}^{(a)} + \tau_{j}^{(a+1)} \right)
\nn \\
&& \hspace{5mm} 
- \Delta_\theta(\T) \left(2\T_{a+2}\tau_{i}^{(a)}\tau_{j}^{(a+1)} 
+ (\T_{a+1} + \T_{a+2})
    (\tau_{i}^{(a)})^2 + (\T_{a+2} + \T_{a})\left(\tau_{j}^{(a+1)}\right)^2
    \right) \biggr] \nn\\
&& \hspace{5mm}-\frac{1}{4} p_{i}^{(a)} \theta^2 p_{j}^{(a+1)} 
   \Delta_\theta (\T) \left[
    \left(\eps_{i}^{(a)}\right)^2 \T_{a+1} 
  + \left(\eps_{j}^{(a+1)}\right)^2 \T_{a}
  + \left(\eps_{i}^{(a)} + \eps_{j}^{(a+1)}\right)^2 \T_{a+2}\right]\nn\\
&& \hspace{5mm}-\frac{1}{2}  p_{i}^{(a)} \theta^2 p_{j}^{(a+1)}
\Delta_\theta (\T) \left[
    \tau_{i}^{(a)}\eps_{j}^{(a+1)}-\eps_{i}^{(a)}\tau_{j}^{(a+1)}
    \right] \Biggr\}\nn\\ 
&&-i\sum_{a=1}^{3}\sum_{ij}^{\N_{a}}p_{i}^{(a)}\wedge p_{j}^{(a)}
    \eps_{j}^{(a)}
-i\sum_{i=1}^{\N_1}\sum_{j=1}^{\N_3}p_{i}^{(1)}\wedge p_{j}^{(3)}
    \nn\\
&&+i{\Delta_\theta}(\T) \sum_{a=1}^{3}\sum_{ij}^{\N_{a}}
   p_{i}^{(a)}\wedge p_{j}^{(a)}(\T_{a+1} + \T_{a+2})
   \tau_{i}^{(a)}\eps_{j}^{(a)}\nn\\
&&-i\Delta_\theta (\T) \sum_{a=1}^{3}\sum_{i=1}^{\N_{a}}\sum_{j=1}^{\N_{a+1}}
    p_{i}^{(a)}\wedge p_{j}^{(a+1)}\left[ \T_{a+2}
  \left(\tau_{i}^{(a)}\eps_{j}^{(a+1)}-\eps_{i}^{(a)}\tau_{j}^{(a+1)} \right)
   -\tau_{i}^{(a)}\tau_{j}^{(a+1)}\right]\nn\\
&&+i\Delta_\theta(\T) \sum_{i=1}^{\N_1}\sum_{j=1}^{\N_3}
   p_{i}^{(1)}\wedge p_{j}^{(3)}(\T_2+\T_3)\tau_{i}^{(1)}
   +i\Delta_\theta (\T) \sum_{i=1}^{\N_3}\sum_{j=1}^{\N_2}
   p_{i}^{(3)}\wedge p_j^{(2)} \T_3\tau_{j}^{(2)}
    \nn\\
&&-i\Delta_\theta (\T) \sum_{i=1}^{\N_3}\sum_{j=1}^{\N_3}
   p_{i}^{(3)}\wedge p_{j}^{(3)}\T_2\tau_{i}^{(3)}
   +\frac{i}{4} \Delta_\theta(\T) 
    \sum_{a=1}^{3}\sum_{i=1}^{\N_{a}}\sum_{j=1}^{\N_{a+1}}
   p_{i}^{(a)}\theta^3 p_{j}^{(a+1)}\eps_{i}^{(a)}\eps_{j}^{(a+1)}
    \nn\\
&& +\frac{1}{2} \Delta_\theta(\T) \sum_{i=1}^{\N_1}\sum_{j=1}^{\N_3}
   p_{i}^{(1)}\circ p_{j}^{(3)}\left[ 
    -(\T_2+\T_3)\eps_{i}^{(1)}-\T_2\eps_{j}^{(3)}+
     \frac{1}{2}(\T_2+\T_3) \right]\nn\\
&&+\frac{1}{2} \Delta_\theta(\T) \sum_{i=1}^{\N_2}\sum_{j=1}^{\N_3}
  p_{i}^{(2)}\circ p_{j}^{(3)}\left[\T_3\eps_{i}^{(2)}- 
  \T_2\eps_{j}^{(3)}+\frac{1}{2}(\T_2+\T_3) \right]\nn\\
&&-\frac{1}{2} \Delta_\theta (\T) \left( \sum_{i=1}^{\N_3}\sum_{j=1}^{\N_2}
   p_{i}^{(3)}\circ p_{j}^{(2)}\tau_{j}^{(2)}
   - \sum_{i=1}^{\N_3}\sum_{j=1}^{\N_3}
   p_{i}^{(3)}\circ p_{j}^{(3)}\tau_{i}^{(3)} \right) \nn\\
&&+\frac{i }{4} \left( \Delta_\theta(\T) \sum_{i=1}^{\N_3}
   \sum_{j=1}^{\N_2}p_{i}^{(3)}\theta^3p_{j}^{(2)}\eps_{j}^{(2)}
+ \sum_{i=1}^{\N_3}
   \sum_{j=1}^{\N_3}p_{i}^{(3)}\theta^3p_{j}^{(3)}\eps_{i}^{(3)}
\right)\ .
\eea
Again, after straightfoward and tedious algebra, we can regroup
terms into cyclically symmetric combinations and obtain the
second exponential and $\Xi$-factor of Eq.(\ref{master2}), where 
the worldline correlators are given by 
\bea
G^{{\bf NP}^{mn}}_{aa}
\left(\tau^{(a)}_i,\tau^{(a)}_j; \eta^{(a)}_i,\eta^{(a)}_j\right)  
&=& g^{mn}G^{\rm sym}_{\theta\, aa}\left(\tau^{(a)}_i,\tau^{(a)}_j\right) 
-2i\eta^{(a)}_j \theta^{mn} \nn\\
&+&2i\Delta_\theta(\T) \theta^{mn}
(\T_{a+1}+\T_{a+2})\tau^{(a)}_i \eta^{(a)}_j \nn\\
&+&\frac{1}{4} \Delta_\theta (\T)(-\theta^2)^{mn}(\T_{a+1} + \T_{a+2})
   \left(\eta_{i}^{(a)} - \eta_{j}^{(a)} \right)^{2}\ ,
\nn
\eea
and 
\bea
G^{{\bf NP}^{mn}}_{aa+1}
\left(\tau^{(a)}_i,\tau^{(a+1)}_j; \eta^{(a)}_i,\eta^{(a+1)}_j\right)  
&=& g^{mn}G^{\rm sym}_{\theta\, aa+1}(\tau^{(a)}_i,\tau^{(a+1)}_j)
+\frac{i}{3} \theta^{mn}\nn\\
&&\hskip-120pt +i \Delta_\theta(\T) \theta^{mn} 
  \left[\, \T_{a+2}\left(\tau_{i}^{(a)} \eta_{j}^{(a+1)}
   - \eta_{i}^{(a)} \tau_{j}^{(a+1)}\right)+\frac{1}{2} \T_{a+1} 
   \tau_{i}^{(a)}+\frac{1}{2} \T_{a}  \tau_{j}^{(a+1)}
    - \tau_{i}^{(a)}\tau_{j}^{(a+1)} \,\right] \nn\\
&&\hskip-120pt+\,\,\frac{1}{4} \Delta_\theta(\T)(-\theta^2)^{mn}
    \left[\left(\eta_{i}^{(a)}+\frac{1}{2}\right)^2 \T_{a+1} 
    + \left(\eta_{j}^{(a+1)}-\frac{1}{2}\right)^2 \T_{a}
   +\left(\eta_{i}^{(a)} + \eta_{j}^{(a+1)}\right)^2 \T_{a+2}\right] \nn\\
&&\hskip-120pt-\frac{1}{2} \Delta_\theta(\T) (-\theta^2)^{mn}
   \left[\left(\eta_{i}^{(a)}+\frac{1}{2}\right) \tau_{j}^{(a+1)}
   - \tau_{i}^{(a)}\left(\eta_{j}^{(a+1)}-\frac{1}{2}\right)\right]\nn\\
&&\hskip-120pt+\frac{i}{4} \Delta_\theta (\T)(\theta^{3})^{mn} 
   \left[\,\left(\eta_{i}^{(a)}+\frac{1}{2}\right)
       \left(\eta_{j}^{(a+1)}-\frac{1}{2}\right)
+\frac{1}{3}\right]\ .
\nn
\eea
while $G^{\rm sym}_{\theta\, ab}$ are defined by replacing $\Delta(\T)$ 
by $\Delta_\theta(\T)$ in $G^{\rm sym}_{ab}$. 
Note that the above expression of the worldline correlators are related to
Eqs.(\ref{npfinala}, \ref{npfinalb}) by the the worldline moduli inversion, 
$\tau^{(a)}\ra \T_a - \tau^{(a)}$.

%
%
\section{Nonplanar Feynman diagrams}\label{ap2}
\setcounter{equation}{0}
\indent


In this appendix, we independently check our worldline computations, 
especially the existence of ${\cal O}(\theta^3)$ effects 
in the nonplanar case. 
While these terms are invisible in the N=2 Green functions and hence 
left unnoticed in earlier work \cite{klp}, 
they begin to be visible for $\N \ge 3$
Green function. For simplicity in illustrating this point, we will consider
a massless $\lambda [\Phi^3]_\star$-theory. 

\vspace{8mm}
\begin{minipage}[htb]{15cm} 
\begin{center}
\input{fig4.pictex}
\end{center}
{\bf Figure 6:} {\sl 
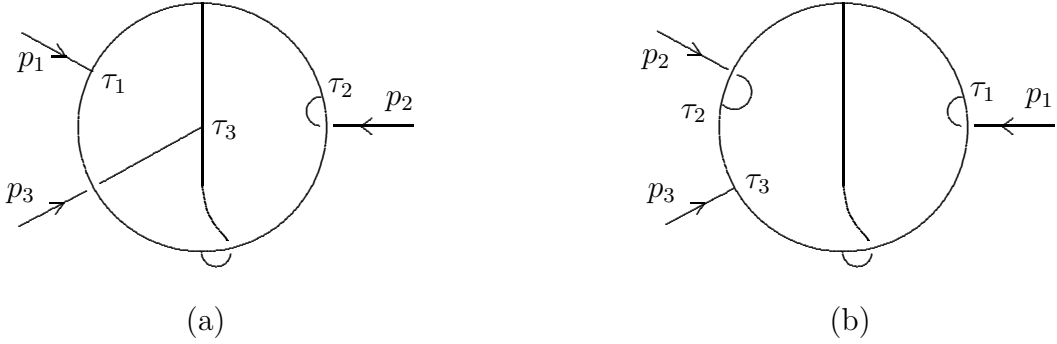
Nonplanar Feynman diagrams computed in Appendix B.} 
\end{minipage}
\vspace{8mm}

The first example is the case of Fig.~6(a), where all $\eta^a$ values are 
equal to $-1/2$, and the overall Filk's phase-factor is equal to 
$e^{{i\over2}p_1\wedge p_2}$: 
\bea
\Lambda_1= e^{{i\over2}p_1\wedge p_2}
\int {\d^d k_1 \over (2 \pi)^d} {\d^d k_2 \over (2 \pi)^d}
{\d^d k_3 \over (2\pi)^{d}}
{\delta(k_1+k_2+k_3)e^{i(k_1+p_1)\wedge p_3}
e^{-ik_2\wedge p_2}e^{i(k_2+p_2)\wedge (k_3+p_3)}
\over k_1^2(k_1+p_1)^2 k_2^2(k_2+p_2)^2 k_3^2(k_3+p_3)^2 } \ .
\nn
\eea
Introducing $d$-dimensional $Y^m$ integrations for the $\delta$-function 
and the parametric integrals such as
\bea
{1\over k_1^2 (k_1+p_1)^2}=\int_0^\infty \d \T_1 \int_0^{\T_1} \d\tau_1 
\exp\Bigl[\,-\tau_1k_1^2-(\T_1-\tau_1)(k_1+p_1)^2\,\Bigr] \ ,
\nn
\eea
we perform $k_a$; $a=1,2,3$, and $Y$ integrals in due course. After 
applying the changes of variables $\tau_a \ra \T_a - \tau_a$, we then have 
\bea
\label{Lam1}
\Lambda_1={1\over(4\pi)^d}
\left(\prod_{a=1}^3 \int_0^\infty \d \T_a \int_0^{\T_a} \d\tau_a\right)
\left(\Delta_\theta(\T)\right)^{d\over2}\,\exp\left[\, 
\tilde{K}^{\mbox{T}}\tilde{A}^{-1}\tilde{K}
+{1\over4}K^{\mbox{T}}A^{-1}K+\tilde{H}_1 \,\right] \ ,
\eea
with 
\bea
&&\tilde{K}={1\over4}(C^{\mbox{T}}A^{-1}K)+{i\over2}Q\ , \\
&&K= i\theta \cdot p_3 + 2\tau_3 p_3 - i{\tau_2\over \T_2}\theta \cdot 
p_2 \ , \\
&&Q= -{\tau_1\over \T_1}p_1 -{\tau_2\over \T_2}p_2 +{i \over 2\T_1}
\theta \cdot p_2
\ , \nn \\
&&\tilde{H}_1 = -p_1^2\left(\tau_1-{\tau^2_1\over \T_1}\right)
-p_2^2\left(\tau_2-{\tau^2_2\over \T_2}\right)
-p_3^2\tau_3 \nn\\
&&\qquad -\, i{\tau_1\over \T_1}p_1\wedge p_2 -{1\over 4 \T_1}p_2\circ p_2
+{i\over2}p_1\wedge p_2 \ ,
\nn
\eea
where $A$ and $C$ are defined in Eq.(\ref{mats}), modified by 
the cyclic permutation $\T_1\ra \T_2 \ra \T_3$. Imposing the momentum 
conservation; $p_1+p_2+p_3=0$, we can rearrange the exponent in Eq.\eq{Lam1} 
in the following form  
\bea
\exp\left[\,   -{i\over2} p_1 \wedge p_2   \,\right]
\exp\left[\, \sum_{a=1}^3 p^m_a p^n_{a+1}
{\cal G}^{mn}_{aa+1}(\tau_a,\tau_{a+1})
\,\right]\ ,
\nn
\eea
where 
\bea
{\cal G}^{mn}_{aa+1}(\tau_a,\tau_{a+1}) &=&
g^{mn}G^{\rm sym}_{\theta\, aa+1}(\tau_a,\tau_{a+1})
-{1 \over 4} \Delta_\theta (\T) (-\theta^2)^{mn}(\T_{a+1}+\T_{a+2} - 2\tau_{a+1}) \nn\\
&-&{i \over 2}\Delta_\theta (\T) \theta^{mn}\Bigl[\, 
T_{a+2}(\tau_a - \tau_{a+1}) +\T_{a+1}\tau_a +\T_{a}\tau_{a+1} 
-2\tau_a\tau_{a+1}\,\Bigr] \nn\\
&+&{i\over3}\left(\, {1 \over 4} \Delta_\theta (\T) 
\theta^3 + \theta\,\right)^{mn}\ .
\nn
\eea
The functions ${\cal G}_{aa+1}$; $a=1,2,3$, coincide with the  
quantity Eq.\eq{npfinalb} when setting $\eta^{(a)}= - {1\over2}$, 
and the extra exponential factor is nothing but $\Xi$ define in Eq.\eq{xi}. 

In a similar fashion, we calculate the second diagram (Fig.~6(b)), where 
the overall Filk's phase-factor amounts to $e^{-{i\over2}p_1\wedge p_2}$ and 
the assignments of external variables are $p^{(1)}_1 = p_2$, 
$p^{(1)}_2 = p_3$, $p^{(2)}_1 = p_1$, $\eta^{(1)}_1 = +{1\over2}$, 
$\eta^{(1)}_2 = -{1\over2}$, $\eta^{(2)}_1 = -{1\over2}$:
\bea
\Lambda_2 &=& e^{-{i\over2}p_1\wedge p_2} 
\int {\d^d k_1 \over (2 \pi)^d} { \d^d k_2 \over (2 \pi)^d}
{\d^d k_3 \over (2\pi)^{d}}
{\delta(k_1+k_2+k_3) e^{ik_1\wedge p_2}
e^{-ik_2\wedge p_1}
e^{i(k_2+p_1)\wedge k_3}
\over k_1^2 (k_1+p_2)^2 (k_1+p_2+p_3)^2 k_3^2 k_2^2 (k_2+p_1)^2 } 
\nn \\
&=&{1\over(4\pi)^d}
\int_0^\infty \d\T_3 
\int_0^\infty \d\T_2 \int_0^{\T_2} \d\tau_2
\int_0^\infty \d\T_1 \int_0^{\T_1} \d\tau_2 \int_0^{\tau_2} \d\tau_1
\, \left(\Delta_\theta (\T)\right)^{d\over2}  \nn\\
&\times& \exp\Bigl[\, 
\tilde{L}^{\mbox{T}}\tilde{A}^{-1}\tilde{L}
+{1\over4}L^{\mbox{T}}A^{-1}L+{\cal H}_1 \,\Bigr] \ ,
\label{Lam2}
\eea
where $A$ is again the permutated one (as well as $C$), and 
\bea
&& \tilde{L}={1\over4}(C^{\mbox{T}}A^{-1}L)+{i\over2}{\cal Q}\ , \nn \\
&& L= i{\tau_1\over \T_2}\theta \cdot p_1 
+ {1\over 2 \T_2}\theta^2 \cdot p_1 \ , \nn \\
&& {\cal Q}= \left({\tau_1\over \T_2}-{i\theta\over 2 \T_2}\right)p_1 
+ \left({\tau_2\over \T_1}+{i\theta\over 2 \T_1}\right)p_2 
+ {\tau_3 \over \T_1}p_3 \ , \nn \\
&&{\cal H}_1 = -p_1^2\left(\tau_1-{\tau^2_1\over \T_2}\right)
-p_2^2\left(\tau_2-{\tau^2_2\over \T_1}\right)
-p_3^2\left(\tau_3-{\tau^2_3\over \T_1}\right)
-2p_2p_3\left(\tau_2-{\tau_2\tau_3\over \T_1}\right) \nn\\
&&\qquad\quad +\, i\left( 1-{\tau_3\over \T_1}\right) p_1\wedge p_2 
-{1\over 4\T_1}p_2\circ p_2 -{1\over 4\T_2}p_1\circ p_1
-{i\over2}p_1\wedge p_2\ .
\nn
\eea

Imposing the overall energy-momentum conservation, and 
changing the moduli variables $\tau_2 \ra \T_2 - \tau_2$; 
$\tau_i \ra \T_1 - \tau_i$; $i=2,3$, we can rearrange the exponential 
in Eq.\eq{Lam2} as 
\bea
\label{exam2}
\exp\left[\, -{i\over2}p_1\wedge p_2 \,\right]
\exp\left[\, {1\over2}\sum_{i,j=2}^3 p_i \cdot
{\cal G}_{11}(\tau_i,\tau_j) \cdot p_j
+\sum_{j=2}^3 p_j \cdot
{\cal G}_{12}(\tau_j,\tau_1) \cdot p_1 \,\right]\ ,
\eea
where 
\bea
&&{\cal G}^{mn}_{11}(\tau_2,\tau_3) =
g^{mn}G^{\rm sym}_{\theta\, 11}(\tau_2,\tau_3)
+i\theta^{mn}(\T_2+\T_3)\tau_2
+{1 \over 4} \Delta_\theta(\T) (-\theta^2)^{mn}(\T_2+\T_3) 
-{i\over 4} \Delta_\theta (\T) \left( \theta^3\right)^{mn},\nn \\
&&{\cal G}^{mn}_{11}(\tau_3,\tau_2) =
g^{mn}G^{\rm sym}_{\theta\, 11}(\tau_2,\tau_3)
-i\theta^{mn}(\T_2+\T_3)\tau_3
+{1 \over 4} \Delta_\theta (\T) (-\theta^2)^{mn}(\T_2+\T_3)
+{i \over 4} \Delta_\theta (\T) (\theta^3)^{mn} \ ,\nn\\
&&{\cal G}^{mn}_{12}(\tau_2,\tau_1) =
g^{mn}G^{\rm sym}_{\theta\, 12}(\tau_2,\tau_1)
-{i \over 2} \Delta_\theta (\T) 
\theta^{mn}\Bigl[\,\T_3(\tau_2+\tau_1)+\T_2\tau_2
+\T_1\tau_1-2\tau_2\tau_1\,\Bigr] \nn\\ 
&&\hskip60pt +\,{i\over3}
\left(\theta - {1 \over 2} \Delta_\theta (\T) \theta^3 \right)^{mn}\ ,\nn \\
&&{\cal G}^{mn}_{12}(\tau_3,\tau_1) =
g^{mn}G^{\rm sym}_{\theta\, 12}(\tau_3,\tau_1)
-{i \over 2} \Delta_\theta (\T) \theta^{mn}
\Bigl[\,\T_3(\tau_3-\tau_1)+\T_2\tau_3+\T_1\tau_1-2\tau_3\tau_1\,\Bigr]
\nn \\
&&\hskip60pt +{1 \over 4} \Delta_\theta (\T) 
(-\theta^2)^{mn}(\T_2+\T_3-2\tau_1) 
+\,{i\over3}
\left(\theta - {1 \over 4} \Delta_\theta (\T) \theta^3 \right)^{mn} \ .
\nn
\eea
These quantities coincide with Eqs.(\ref{npfinala}, \ref{npfinalb}) for the 
present values of $\eta^{(a)}$ mentioned above. The extra factor 
appearing in Eq.\eq{exam2} is nothing but $\Xi$ defined by Eq.\eq{xi}.  

\newpage


\end{document}